\documentstyle[11pt,amssymb,epsf]{article}
\def\ben{\begin{equation}}
\def\een{\end{equation}}

\let\a=\alpha    \let\e=\epsilon
    
  \let\n=\nu

\let\C=\Chi

\def\nn{\nonumber} \def\bd{\begin{document}} \def\ed{\end{document}}
\def\ds{\documentstyle} \let\fr=\frac \let\bl=\bigl \let\br=\bigr
\let\Br=\Bigr \let\Bl=\Bigl
\let\bm=\bibitem
\let\na=\nabla
\let\pa=\partial \let\ov=\overline
\newcommand{\be}{\begin{equation}}
\newcommand{\ee}{\end{equation}}
\def\ba{\begin{array}}
\def\ea{\end{array}}
\def\ft#1#2{{\textstyle{{\scriptstyle #1}\over {\scriptstyle #2}}}}
\def\fft#1#2{{#1 \over #2}}
\def\del{\partial}
\def\vp{\varphi}
\def\sst#1{{\scriptscriptstyle #1}}
\def\oneone{\rlap 1\mkern4mu{\rm l}}
\def\td{\tilde}
\def\wtd{\widetilde}
\def\ie{\rm i.e.\ }
\def\dalemb#1#2{{\vbox{\hrule height .#2pt
        \hbox{\vrule width.#2pt height#1pt \kern#1pt
                \vrule width.#2pt}
        \hrule height.#2pt}}}
\def\square{\mathord{\dalemb{6.8}{7}\hbox{\hskip1pt}}}
\newcommand{\ho}[1]{$\, ^{#1}$}
\newcommand{\hoch}[1]{$\, ^{#1}$}
\newcommand{\bea}{\begin{eqnarray}}
\newcommand{\eea}{\end{eqnarray}}
\newcommand{\ra}{\rightarrow}
\newcommand{\lra}{\longrightarrow}
\newcommand{\Lra}{\Leftrightarrow}
\newcommand{\ap}{\alpha^\prime}
\newcommand{\bp}{\tilde \beta^\prime}
\newcommand{\tr}{{\rm tr} }
\newcommand{\Tr}{{\rm Tr} }
\def\0{{\sst{(0)}}}
\def\1{{\sst{(1)}}}
\def\2{{\sst{(2)}}}
\def\3{{\sst{(3)}}}
\def\4{{\sst{(4)}}}
\def\5{{\sst{(5)}}}
\def\6{{\sst{(6)}}}
\def\7{{\sst{(7)}}}
\def\8{{\sst{(8)}}}
\def\n{{\sst{(n)}}}
\def\cA{{{\cal A}}}
\def\cF{{{\cal F}}}
\def\tV{\widetilde V}
\def\tW{\widetilde W}
\def\tH{\widetilde H}
\def\tE{\widetilde E}
\def\tF{\widetilde F}
\def\tA{\widetilde A}
\def\im{{{\rm i}}}
\def\tY{{{\wtd Y}}}
\def\ep{{\epsilon}}
\def\vep{{\varepsilon}}
\def\R{\rlap{\rm I}\mkern3mu{\rm R}}
\def\bD{{{\bar D}}}

\def\R{\rlap{\rm I}\mkern3mu{\rm R}}
\def\bD{{{\bar D}}}
\def\R{{{\Bbb R}}}
\def\C{{{\Bbb C}}}
\def\H{{{\Bbb H}}}
\def\CP{{{\Bbb C}{\Bbb P}}}
\def\RP{{{\Bbb R}{\Bbb P}}}
\def\Z{{{\Bbb Z}}}
\def\bA{{{\Bbb A}}}
\def\bB{{{\Bbb B}}}
\def\bC{{{\Bbb C}}}
\def\bR{{{\Bbb R}}}
\def\bD{{{\Bbb D}}}
\def\bE{{{\Bbb E}}}
\def\bZ{{{\Bbb Z}}}
\def\Re{{{\frak{Re}}}}
\def\Im{{{\frak{Im}}}}
\def\cosec{{\,\hbox{cosec}\,}}
\def\Gm{{\Gamma_{\!\! -}}}
\def\Gp{{\Gamma_{\!\! +}}}
\def\stan{{standard }}
\def\nonstan{{supernumerary }}

\def\cosech{{\hbox{cosech}}}

\def\etcyc{{\hbox{and cyclic}}}

\newcommand{\tamphys}{\it Center for Theoretical Physics,
Texas A\&M University, College Station, TX 77843, USA}
\newcommand{\umich}{\it Michigan Center for Theoretical Physics,
University of Michigan\\ Ann Arbor, MI 48109, USA}
\newcommand{\upenn}{\it Department of Physics and Astronomy,
University of Pennsylvania\\ Philadelphia,  PA 19104, USA}
\newcommand{\SISSA}{\it  SISSA-ISAS and INFN, Sezione di Trieste\\
Via Beirut 2-4, I-34013, Trieste, Italy}

\newcommand{\newton}{\it Isaac Newton Institute for Mathematical Sciences,\\
20 Clarkson Road,  University of Cambridge,
Cambridge CB3 0EH, UK}

\newcommand{\ihp}{\it Institut Henri Poincar\'e\\
  11 rue Pierre et Marie Curie, F 75231 Paris Cedex 05}

\newcommand{\damtp}{\it DAMTP, Centre for Mathematical Sciences,
 Cambridge University\\ Wilberforce Road, Cambridge CB3 OWA, UK}
\newcommand{\itp}{\it Institute for Theoretical Physics, University of
California\\ Santa Barbara, CA 93106, USA}

\newcommand{\auth}{M. Cveti\v{c}\hoch{\dagger\flat}, 
G.W. Gibbons\hoch{\sharp},
H. L\"u\hoch{\star} and C.N. Pope\hoch{\ddagger\flat}}

\begin{document}
\begin{flushright}
\hfill{DAMTP-2002-65}\ \ \ {CTP TAMU-13/02}\ \ \ {UPR-1002-T}\ \ \
{MCTP-02-28} \ \ \  {NI02018-MTH}\\
{June 2002}\ \ \
{hep-th/0206151}
\end{flushright}

%\vspace{15pt}

\begin{center}
{ \Large {\bf Bianchi IX Self-dual Einstein Metrics and  
Singular $G_2$ Manifolds}}

\vspace{5pt}
\auth

\vspace{3pt}
{\hoch{\dagger}\upenn}

\vspace{3pt}

\vspace{3pt}
{\hoch{\sharp}\damtp}

\vspace{3pt}
{\hoch{\star}\umich}

\vspace{3pt}
{\hoch{\ddagger}\tamphys}

\vspace{3pt}
{\hoch{\flat}\newton}

\vspace{3pt}

\underline{ABSTRACT}
\end{center}

    We construct explicit cohomogeneity two metrics of $G_2$ holonomy,
which are foliated by twistor spaces.  The twistor spaces are $S^2$
bundles over four-dimensional Bianchi IX Einstein metrics with
self-dual (or anti-self-dual) Weyl tensor.  Generically the 4-metric
is of triaxial Bianchi IX type, with $SU(2)$ isometry.  We derive the
first-order differential equations for the metric coefficients, and
obtain the corresponding superpotential governing the equations of
motion, in the general triaxial Bianchi IX case.  In general our
metrics have singularities, which are of orbifold or cosmic-string
type. For the special case of biaxial Bianchi IX metrics, we give a
complete analysis their local and global properties, and the
singularities. In the triaxial case we find that a system of equations
written down by Tod and Hitchin satisfies our first-order equations.
The converse is not always true. A discussion is given of the possible
implications of the singularity structure of these spaces for M-theory
dynamics.

\pagebreak
\setcounter{page}{1}

\tableofcontents
\addtocontents{toc}{\protect\setcounter{tocdepth}{3}}
\vfill\eject

\section{Introduction}

   Concrete non-singular examples of seven-dimensional metrics with
$G_2$ holonomy have been known only since about 1989.  The original
construction involved making an ansatz for metrics of cohomogeneity
one, where the six-dimensional principal orbits were $S^3\times S^3$,
or else the twistor spaces of $S^4$ or $\CP^2$
\cite{brysal,gibpagpop}.  The twistor space is a 2-sphere bundle over
the $S^4$ or $\CP^2$ base, with an $SU(2)$ or $SO(3)$ structure group
associated to the chiral spin (or spin$^c$) bundle of the base.  The
local construction can be carried out for any base space $M_4$
equipped with an Einstein metric and for which the Weyl tensor is
self-dual or anti-self-dual \cite{brysal,gibpagpop}.\footnote{Such
metrics are generally referred to as ``self-dual Einstein,'' and
unless the context makes it necessary in order to avoid confusion, we
shall often use this term regardless of whether the Weyl tensor is actually
self-dual or anti-self-dual.}  By a theorem of Hitchin, the only
nonsingular such examples with positive Ricci tensor (which implies
$M_4$ is compact) occur when $M_4$ is $S^4$ or $\CP^2$
\cite{hitchbess}.

   The current interest in $G_2$ manifolds in M-theory has been
motivated in part by the role that they can play in compactifying to
four dimensions, analogous to the compactification of ten-dimensional
string theory on Calabi-Yau 6-manifolds.  Unlike the latter, where
non-singular Calabi-Yau manifolds can naturally give rise to chiral
${\cal N}=1$ theories in four dimensions starting from the heterotic
string in $D=10$, non-singular $G_2$ compactifications of M-theory
would necessarily give abelian non-chiral ${\cal N}=1$ theories in
four dimensions.  To get non-abelian chiral theories from M-theory,
one needs to consider compactifications on singular $G_2$
manifolds. One explicit realisation of such an M-theory
compactification has an interpretation as an $S^1$ lift of Type IIA
theory (compactified on an orientifold) with intersecting D6-branes
and O6 orientifold planes \cite{CSUII}. Non-Abelian gauge fields arise
at the locations of coincident branes, and chiral matter arises at the
intersections of D6-branes.  The $S^1$ lift of such configurations
results in singular $G_2$ holonomy metrics in M-theory.  Co-dimension
four ADE-type singularities are associated with the location of the
coincident D6-branes, and co-dimension seven singularities are
associated with the location of the intersection of two D6-branes in
Type IIA theory \cite{CSUII,AW,Witten,AcW,CSUIII}.

   Further analyses of co-dimension seven singularities of the
$G_2$ holonomy spaces, leading to chiral matter, were given in
\cite{Witten,AcW,CSUIII} and the subsequent work
\cite{Roiban,behrndt,LazarI,bb}.  It is expected that there exist wide
classes of 7-manifolds with $G_2$ holonomy and the singularity
structure that again would yield non-Abelian ${\cal N}=1$
supersymmetric four-dimensional theories with chiral matter, and in
particular the explicit construction of such metrics would provide a
starting point for further studies of chiral M-theory dynamics.

  Much research on finding new non-singular $G_2$ manifolds has been
carried out in recent times (see, for example,
\cite{cglp10,bggg,cglp11,brand,cglp12,zaslow,cglp14} and references
therein).  In view of their potential phenomenological interest, it is
appropriate also to investigate examples of singular $G_2$ manifolds.
Typically, these singularities should be of co-dimension seven, and
they should be of the relatively mild orbifold type \cite{AW,Witten},
where the curvature is bounded everywhere except for delta-function
contributions.

    One way to obtain singular $G_2$ holonomy spaces is by returning
to the original $G_2$ construction in \cite{brysal,gibpagpop}, with
principal orbits that are $S^2$ bundles over self-dual Einstein
four-dimensional manifolds $M_4$ (forming the base of the twistor
space), but with $M_4$ now chosen to be neither the $S^4$ nor the
$\CP^2$ non-singular examples. Instead, one can choose $M_4$ to be a
self-dual Einstein space with orbifold-type singularities.  Some
investigations of the $G_2$ metrics that result from such a
construction have already been carried out \cite{behrndt}. In this
paper we pursue the analysis further, by considering more general
possibilities for the base space $M_4$.  Since the procedure for
obtaining the $G_2$ metric from a given self-dual Einstein base space
$M_4$ is well established \cite{brysal,gibpagpop}, much of the paper
will concentrate on the details of the self-dual Einstein metrics
themselves.
  
   There exists a large mathematical literature on self-dual Einstein
metrics (sometimes called quaternionic K\"ahler).  The focus of our
study in this paper will be on self-dual Einstein metrics of the
triaxial Bianchi IX type, where there is an $SU(2)$ isometry that acts
transitively on 3-dimensional orbits that are (locally) $S^3$.  Quite
a lot is known about this case \cite{todrev,hitchin,hitchin2}, but we
believe that our results go beyond what is in the existing literature,
and that our viewpoint, derived as it is from the associated $G_2$
metric, is novel.  In particular, we shall derive the general
first-order equations for these metrics and analyse their local and
global structure. For the special case of biaxial Bianchi IX metrics,
we provide a complete analysis.  In the triaxial case, we compare our
analysis with that of Tod \cite{tod} and Hitchin
\cite{hitchin,hitchin2}, and analyse some of the explicitly-known
solutions.  Some implications for M-theory of these $G_2$ holonomy
metrics are also discussed.

\section{Asymptotically-conical $G_2$ metrics}\label{acg2sec}

\subsection{$G_2$ holonomy of $R^3$ bundles over self-dual 
            Einstein 4-metrics}\label{g2sdweylsec}

   The metrics of $G_2$ holonomy that have
twistor-space orbits take the form \cite{brysal,gibpagpop}
%%%%%
\be
ds_7^2 = 4\Big(1-\fft{1}{r^4}\Big)^{-1}\, dr^2 + 
     r^2 \, \Big(1-\fft{1}{r^4}\Big)\, (D\mu_i)^2 + 2 r^2\, ds_4^2\,,
\label{g2orig}
\ee
%%%%%
where $\mu_i\, \mu_i=1$.  The covariant exterior derivative is defined
by $D\mu_i\equiv d\mu_i + \ep_{ijk}\, A^j\, \mu_k$, and the metric 
$ds_4^2$ is required to be Einstein, with $R_{ab}=\Lambda\, g_{ab}$
(with $\Lambda$ taken to be normalised to $\Lambda=3$ in (\ref{g2orig}).
The Yang-Mills fields have the defining property that $D\, J^i=0$,
where the quaternionic K\"ahler forms $J^i$ on the base space M$_4$ 
have a definite duality, and satisfy
%%%%%
\be
J^i_{ab}\,J^j_{bc} = -\delta_{ac}\, \delta^{ij} + \ep_{ijk}\, 
J^k_{ac}\,,
\ee
%%%%% 
where the gauge-covariant exterior derivative is defined by
%%%%%
\be
D\, J^i_{ab} \equiv \nabla\, J^i_{ab} + \ep_{ijk}\, A^j\, J^k_{ab}\,,
\ee
%%%%%
with
%%%%%
\be
\nabla\, J^i_{ab}\equiv d\, J^i_{ab} + \omega_{ac}\, J^i_{cb} +
\omega_{bc}\, J^i_{ac}\,.
\ee
%%%%%
The integrability condition $D^2\, J^i_{ab}=0$ has, as a particular
consequence,
%%%%%
\be
F^i_{ab} = \ft12 J^i_{cd}\, R_{abcd}\,,\label{fjr}
\ee
%%%%%%
where $F^i\equiv  dA^i +\ft12 \ep_{ijk}\, A^j\wedge A^k$.  
We furthermore require that the Yang-Mills fields $F^i$ be proportional
to the quaternionic K\"ahler forms.  We shall take $J^i_{ab}$ to 
be self-dual, in which case we have the identity
%%%%%%
\be
J^i_{ab}\, J^i_{cd} = \delta_{ac}\, \delta_{bd} - \delta_{ad}\, 
\delta_{bd} + \ep_{abcd}\,.
\ee
%%%%%%
From this and equation (\ref{fjr}), it can be seen that if the Weyl tensor
%%%%%
\be
C_{abcd}\equiv R_{abcd} - \ft13\Lambda\, (\delta_{ac}\, \delta_{bd} -
\delta_{ad}\, \delta_{bc}) 
\ee
%%%%%%
of the Einstein metric $ds_4^2$ is anti-self-dual, then we shall have 
%%%%%%
\be
F^i= \ft13\Lambda\, J^i\,.
\ee
%%%%%

   We can change variables to a set of coordinates $u_i$ on $\R^3$,
which are unconstrained, by taking $u_i=\rho\, \mu_i$, and letting
$\ft13\Lambda\, \rho^2=r^4-1$, leading to the expression
%%%%%%
\be
ds_7^2 = \fft{(Du_i)^2}{\sqrt{1+ \ft13\Lambda\, \rho^2}} + 2\sqrt{
1+\ft13\Lambda\, \rho^2}\, ds_4^2\,,\label{g2met}
\ee
%%%%%
where $\rho$ means $\sqrt{u_i\, u_i}$, $D\, u_i = d\, u_i + \ep_{ijk}\, 
A^j\, u_k$, and we have rescaled so that
$ds_4^2$ has cosmological constant $\Lambda$.

   The $G_2$ holonomy is easily established by noting that we may take
the associative 3-form to be, reverting to $\Lambda=3$ for convenience,
%%%%%
\be
\Phi_\3 = \ft16 (1+\rho^2)^{-3/4}\, \ep_{ijk}\, Du_i\, Du_j\, 
Du_k + 2(1+\rho^2)^{1/4}\, Du_i \wedge J^i\,.
\label{phi}
\ee
%%%%%
The dual of $\Phi_\3$ in the metric (\ref{g2met}) is therefore
%%%%%
\be
{*\Phi_\3} = 4(1+\rho^2)\, 
\Omega_\4 + \ep_{ijk}\, Du_i\wedge Du_j\wedge
J^k\,,\label{starphi}
\ee
%%%%%
where $\Omega_\4$ is the volume form of $ds_4^2$, which can also be
written as $\Omega_\4=\ft12 J^1\wedge J^1 = \ft12 J^2\wedge J^2 =
\ft12 J^3\wedge J^3$.  From the identity $D^2\, u_i = \ep_{ijk}\,
F^j\, u_k$, one easily sees that $\Phi_\3$ is closed and co-closed.

\subsection{Nearly-K\"ahler geometry and $G_2$ holonomy}

   If we go to the asymptotic region, where $\rho\longrightarrow\infty$, 
we get the metric on the cone over the twistor space of M$_4$,
%%%%
\be
ds_7^2 = \fft1{\rho}\,(Du_i)^2 + 2\,  \rho\, ds_4^2\,.\label{g2conemet}
\ee
%%%%% 
Defining $\rho=\ft14 r^2$, this becomes
%%%%%
\be
ds_7^2 = dr^2 + r^2\, ds_6^2\,,
\ee
%%%%%
and so if $ds_7^2$ has $G_2$ holonomy then
%%%%%
\be
ds_6^2 = \ft14 (D\mu_i)^2 + \ft12  ds_4^2\label{twistor}
\ee
%%%%%
is the nearly-K\"ahler metric on the twistor space of M$_4$. 
The associative 3-form becomes 
%%%%%
\be
\Phi_\3 = \ft16 \rho^{-3/2}\,  
\ep_{ijk}\, Du_i\, Du_j\,Du_k
+ 2 \rho^{1/2}\,  Du_i \wedge J^i\,,
\label{acphi}
\ee
%%%%%%
and its Hodge dual is
%%%%%
\be
{*\Phi_\3} = 4 \rho^2\, 
\Omega_\4 + \ep_{ijk}\, Du_i\wedge Du_j\wedge J^k\,.
\label{acstarphi}
\ee
%%%%%
The conditions of closure and co-closure of $\Phi_\3$ therefore imply
that $ds_6^2$ in (\ref{twistor}) is nearly-K\"ahler.  

   The definition of a nearly-K\"ahler metric $ds_6^2$ is that the cone
over $ds_6^2$, namely
%%%%%
\be
d\hat s_7^2 = dr^2 + r^2\, ds_6^2\,\label{7to6met}
\ee
%%%%%
has $G_2$ holonomy.  A more suggestive, but equivalent, terminology
for $ds_6^2$ is therefore that it has {\it weak $SU(3)$ holonomy}; we
discuss this briefly below.
 
   If the cone metric $d\hat s_7^2$ has $G_2$ holonomy, it follows that
the associative 3-form $\Phi_\3$, which may be written as
%%%%%
\be
\Phi_\3 = r^2\, dr\wedge J_\2 + r^3\, \rho_\3\,,
\ee
%%%%%
must be closed and co-closed.  This has the consequences
%%%%%
\be
dJ_\2 = 3\rho_\3\,,\qquad d\td\rho_\3 + 2J_\2\wedge J_\2=0\,,\label{jrho}
\ee
%%%%%
where $\td\rho_\3\equiv {*_6 \rho_\3}$.  Immediate further
consequences of these equations are $J_\2\wedge \rho_\3=0$ and
$d\rho_\3=0$.  The associativity relation
%%%%%
\be
\Phi_{ABE}\, \Phi_{CDE} = \delta_{AC}\, \delta_{BD} - \delta_{AD}\, 
\delta_{BC} + \ft16\ep_{ABCDEFG}\, \Phi_{EFG}
\ee
%%%%%
for the 3-form $\Phi_\3$ has the consequences that 
%%%%%
\be
J_{ab}\, J_{bc} = -\delta_{ac}\,,\qquad
J_{ad}\, \rho_{abd} = \td\rho_{abc}\,.
\ee
%%%%%
This means that $J$ defines an almost complex structure in
$ds_6^2$, and that with respect to $J$, $\psi_\3\equiv \rho+ \im\,
\hat\rho$ is a holomorphic 3-form of type (3,0).

   The relation to weak $SU(3)$ holonomy can be made more explicit by
considering the covariantly-constant spinor $\hat\eta$ that exists in
the $G_2$ metric (\ref{7to6met}).  In the natural orthonormal basis
$\hat e^0=dr$, $\hat e^a= r\, e^a$, one finds that the covariant
exterior derivative $\hat\nabla = d+\ft14 \hat\omega_{AB}\,
\Gamma^{AB}$ is given by $\hat\nabla=\nabla -\ft12 \Gamma_{0a} e^a +
dr\, \fft{\del}{\del r}$.  If $ds_6^2$ has weak $SU(3)$ holonomy then
it admits Killing spinors $\eta_\pm$ satisfying $D^\pm_a\, \eta_\pm
\equiv (\nabla_a \pm \ft12 \Gamma_{0a})\, \eta_\pm=0$, for which the
integrability condition is $[D^\pm_a, D^\pm_b]= \ft14 C_{abcd}\,
\Gamma^{cd}$.  This admits solutions $\eta_\pm$ if $C_{abcd}\,
\Gamma^{cd}$ generates the $SU(3)$ subgroup of the tangent-space group
$SO(6)\sim SU(4)$.  The two spinors are related by $\eta^*_\pm =
\eta_\mp$, and $\Gamma_0\, \eta_\pm =-\eta_\mp$.  The covariantly
constant spinor in the $G_2$ metric $d\hat s_7^2$ is given by $\hat
\eta = \eta_-$.

   In terms of the Killing spinors $\eta_\pm$ in $ds_6^2$, the almost 
complex structure $J_\2$ and the 3-form $\rho_\3$ are given by
%%%%%
\be
J_{ab} = \im\, \eta^\dagger_+\, \Gamma_{ab}\, \eta_-\,,\qquad
\rho_{abc} = \im\, \eta^\dagger_+\, \Gamma_{0abc}\, \eta_-\,.
\ee
%%%%%
From the Killing spinor equations $D^\pm_a\, \eta_\pm=0$ one can now
easily derive the equations
%%%%%
\be
\nabla_a\, J_{bc} =\rho_{abc}\,, \qquad
\nabla_a\, \td\rho_{bcd} = -3 J_{[ab}\, 
J_{cd]}\,.
\ee
%%%%%
These equations, which in particular imply (\ref{jrho}), characterise
nearly-K\"ahler metrics.  Note that by symmetrising the first equation
on $a$ and $b$, we obtain the equation for a Yano Killing tensor,
$\nabla_{(a}\, J_{b)c}=0$ \cite{yano}.  Thus the nearly K\"ahler
6-manifolds constructed in this paper provide new examples of a
supersymmetric quantum mechanical systems with hidden symmetries
\cite{giriho,azizma}.  In fact, because $J_{ab}$ is an almost complex
structure, the associated symmetric Staeckel Killing tensor is given
by $J_{ab}\, J^b{}_c=-g_{ac}$, and hence is trivial in this case.

\section{$G_2$ holonomy equations for Bianchi IX base}\label{foeqsec}

    We now apply the formalism of section \ref{g2sdweylsec}, with the
four-dimensional base metric taken to be of the triaxial Bianchi IX
form:
%%%%%
\be
ds_4^2 = dt^2 + a_i^2\, \sigma_i^2\,.\label{bianchi9}
\ee
%%%%%
The self-dual $SU(2)$ Yang-Mills connection is
%%%%%
\be
A^i = -\omega_{0i} - \ft12\ep_{ijk}\, \omega_{jk}\,,\label{sdym}
\ee
%%%%%
where the spin connection of $ds_4^2$, in the vielbein basis
$e^0=dt$, $e^i=a_i\, \sigma_i$, is given by
%%%%%
\be
\omega_{01} = \beta_1\, e^1\,,\qquad 
\omega_{23}= \gamma_1\, e^1\,,\label{spincon}
\ee
%%%%%
and cyclically, with
%%%%%
\be
\beta_1 \equiv -\fft{\dot a_1}{a_1}\,,\qquad
\gamma_1\equiv \fft{a_1^2- a_2^2 -a_3^2}{2 a_1\, a_2\, a_3}\,,
\label{bega}
\ee
%%%%%
and cyclically.  Since the Yang-Mills potentials are expressed in
terms of the left-invariant 1-forms $\sigma_i$, 
%%%%%
\be
A^i= -a_i\, (\beta_i+\gamma_i)\, \sigma_i\,,\label{asig} 
\ee
%%%%%
the field strengths are necessarily $SU(2)$ invariant, and are given
by
%%%%%
\be
F^i = -\Theta_{0i} -\ep_{ijk}\, \Theta_{jk}\,.\label{fexp}
\ee
%%%%%

   By imposing the closure and co-closure of $\Phi_\3$ given by
(\ref{phi}) (or equivalently, and more simply, (\ref{acphi})), we find
that the first-order equations for $a_i$ such that the 7-manifold has
$G_2$ holonomy are then given by
%%%%%
\bea
&&\dot a_1 -\dot a_2\, \dot a_3 + 
\Big(\fft{a_3^2 - a_1^2 - a_2^2}{2a_1\, a_2}\Big)\,\dot a_2 +
\Big(\fft{a_2^2-a_1^2-a_3^2}{2a_1\, a_3}\Big)\, \dot a_3 \nn\\
&&\qquad\qquad\qquad+\fft{a_2^4+a_3^4-3a_1^4 +2(a_1^2\, a_2^2 + 
a_1^2\, a_3^2 -
a_2^2\, a_3^2- \ft23\Lambda\, 
       a_1^2\, a_2^2\, a_3^2\,)}{4 a_1^2\, a_2\, a_3}=0\,,
\label{foeq}
\eea
%%%%%
together with the two equations obtained by cyclic permutation of the
subscripts 1, 2 and 3.  Note that we have restored the cosmological 
constant $\Lambda$, so that $ds_4^2$ satisfies $R_{ab}=\Lambda\, g_{ab}$.
It is straightforward to see that after using the first-order
equations, (\ref{fexp}) becomes
%%%%%
\be
F^i= -\ft13 \Lambda\, (e^0\wedge e^i + \ft12\ep_{ijk}\, e^j\wedge
e^k) = \ft13\Lambda\, J^i\,.
\ee
%%%%%
   
    We saw in in section \ref{acg2sec} that the conditions for
$ds_7^2$ in (\ref{g2met}) to have $G_2$ holonomy should be equivalent
to the conditions for $ds_4^2$ to have (anti)-self-dual Weyl tensor.  In
fact another way to derive the first-order equations (\ref{foeq}) is as 
follows.  We define the family of tensors 
%%%%%
\be
X_{abcd}\equiv R_{abcd} - \kappa\, (g_{ac}\, g_{bd} - g_{ad}\, g_{bc}\,,
\label{xdef}
\ee
%%%%%
where $\kappa$ is an as-yet unspecified constant parameter.  If we
now require that $X_{abcd}$ be anti-self-dual, we obtain the 
equation
%%%%%
\be
^*R_{abcd} +R_{abcd} -\kappa\, \ep_{abcd} -\kappa\, (g_{ac}\, g_{bd} 
-g_{ad}\, g_{bd})=0\,.
\ee
%%%%%
Contraction with $g^{bd}$ gives $R_{ac} = 3\kappa\, g_{ac}$.  It then
follows that $X_{abcd}$ is the Weyl-tensor of an Einstein metric with
scalar curvature $12\kappa$, and moreover that the Einstein manifold
has anti-self-dual Weyl tensor.  We find that the equations 
%%%%%
\be
X_{0123}= -X_{2323}\,,\qquad X_{0231}=-X_{3131}\,,\qquad X_{0312}=-X_{1212}
\ee
%%%%%
give precisely (\ref{foeq}), and that the remaining anti-self-duality
equations for $X_{abcd}$, \ie 
%%%%%
\be
X_{0101}=-X_{0123}\,,\qquad X_{0202}=-X_{0231}\,,\qquad 
X_{0303}=-X_{0312}\,,
\ee
%%%%%
give second-order equations that are nothing but the derivatives of
(\ref{foeq}).

   One could in principle solve (\ref{foeq}) for the
$\dot a_i$ themselves, but this involves finding the roots of a
quintic equation.  It is, nevertheless, useful to present the
first-order equations in a factorised form.  Solving two of the equations
(\ref{foeq}) for $\dot a_2$ and $\dot a_3$, and substituting into the
third, we get
%%%%%
\bea
&&\Big(\dot a_1 - \fft{a_1^2 -(a_2+a_3)^2}{2 a_2\, a_3}\Big)^2\, 
\Big(\dot a_1 - \fft{a_1^2 -(a_2-a_3)^2}{2 a_2\, a_3}\Big)^2
\Big(\dot a_1 - \fft{a_1^2 -a_2^2-a_3^2}{2 a_2\, a_3}-\ft13 \Lambda\, 
 a_2\, a_3\Big)\nn\\
&&-\ft19\Lambda^2\, a_1^2\, a_2\, a_3 (2 a_2\, a_3\, \dot a_1 + 3a_2^2 +
   a_3^2 -a_1^2)(2 a_2\, a_3\, \dot a_1 + a_2^2 +
   3a_3^2 -a_1^2) =0\,.\label{a1eq}
\eea
%%%%% 
Of course the two equations following by cyclic permutation hold too, but
it would be misleading to think of these three as the equations for the
$a_i$, since one should not solve them independently.  Rather, we
can view (\ref{a1eq}) itself as the equation for $\dot a_1$, and then 
substitute this solution back into the cyclic set defined by (\ref{foeq}) 
in order to obtain the equations for $\dot a_2$ and $\dot a_3$.  

   It is interesting to observe that in the limit when
$\Lambda\longrightarrow 0$, then from (\ref{a1eq}) and (\ref{foeq}) we 
can see that we get either the ``Atiyah-Hitchin'' \cite{atihit} first-order 
system\footnote{Or an equivalent one with sign reversals of certain
of the $a_i$ functions}
%%%%%
\be
\dot a_1 = \fft{a_1^2 - (a_2+a_3)^2}{2 a_2\, a_3}\,,\qquad
\hbox{and cyclic}\,,\label{ahsys}
\ee
%%%%%
or the ``BGPP'' \cite{begipapo} system
%%%%%
\be
\dot a_1 = \fft{a_1^2 - a_2^2 -a_3^2}{2 a_2\, a_3}\,,\qquad
\hbox{and cyclic}\,.\label{bgppsys}
\ee
%%%%%
The equations (\ref{ahsys}) admit the Atiyah-Hitchin \cite{atihit} and
self-dual Taub-NUT \cite{hawking} metrics as particular solutions,
whilst the equations (\ref{bgppsys}) admit the BGPP \cite{begipapo}
and Eguchi-Hanson \cite{eguhan} metrics as solutions.

   It is often more convenient to recast first-order equations such
as (\ref{foeq}) into a form where the metric functions $\a_i\equiv
a_i^2$ themselves appear without square roots.  This can be achieved
by introducing a new radial variable $\rho$, defined by $dt=a_1\,
a_2\, a_3\, d\rho$.  We then find that (\ref{foeq}) becomes
%%%%%
\bea
&&2 \fft{d\a_1}{d\rho} - \fft1{\a_2\,\a_3}\, \fft{d\a_2}{d\rho}\, 
\fft{d\a_3}{d\rho} + \fft{(\a_3-\a_1-\a_2)}{\a_2}\, \fft{d\a_2}{d\rho} +
\fft{(\a_2-\a_3-\a_1)}{\a_3}\, \fft{d\a_3}{d\rho} \nn\\
&&+ \a_2^2+\a_3^2-3\a_1^2 + 2\a_1\,(\a_2+\a_3) - 2\a_2\, \a_3 
   -\ft{4}{3}\, \Lambda \a_1\, \a_2\, \a_3=0\,,\label{foeq2}
\eea
%%%%%
and cyclically.  Note also that in terms of the $\beta_i$ and $\gamma_i$
coefficients defined in (\ref{spincon}) and (\ref{bega}), the
first-order equations (\ref{foeq}) can be written as
%%%%%
\be
(\beta_1 + \gamma_1)\, (\gamma_2+\gamma_3) = (\beta_2+\gamma_2)\, 
(\beta_3+\gamma_3) +\ft13 \Lambda\,,
\ee
%%%%%
and cyclically.

   If we consider the specialisation where all three metric functions
$a_i$ are set equal, $a_i=a$, the first-order system (\ref{foeq}) reduces
to
%%%%%
\be
\dot a^2 + \ft13\Lambda\, a^2 = \ft14\,.\label{uniaxialfo}
\ee
%%%%%
This gives 
%%%%%
\be
ds_4^2 = dt^2 + \fft{3}{4\Lambda}\, \sin^2( \sqrt{\ft13\Lambda}\,\, t)\, 
\sigma_i^2\,.
\ee
%%%%%
The metric extends to a complete non-singular metric on $S^4$ if
$\Lambda>0$, and to the hyperbolic space $H^4$ if $\Lambda<0$.

    The specialisation to biaxial metrics, where two of the metric
functions are set equal, is considerably more complicated.  We shall 
study this in detail in the next section.

\section{Biaxial anti-self-dual Bianchi IX metrics}

   In this section we shall specialise to the biaxial case, 
setting $a_2=a_1$.  The first-order equations (\ref{foeq}) reduce to
%%%%%
\bea
\dot a_3 &=& \dot a_1^2 + \fft{\dot a_1\, a_3}{a_1} -1 + 
                 \fft{3 a_3^2}{4 a_1^2} + \ft13\Lambda\, a_1^2\,,\nn\\
0&=&(2 a_1\, \dot a_1 + a_3)(2a_1\, \dot a_1 + a_3 -2a_1)(2a_1\, \dot a_1 
   + a_3 + 2a_1) + \ft16 \Lambda\, a_1 (2 a_1\, \dot a_1 + 3a_3)\,.
\label{biaxialfo}
\eea
%%%%%
It is easy to see that if we take the limit where $\Lambda$ goes to
zero, the cubic equation for $\dot a_1$ has roots giving
%%%%%
\be
\dot a_1 = -\fft{a_3}{2a_1} +1\,,\quad \hbox{or}\quad
\dot a_1 = -\fft{a_3}{2a_1} -1\,,\quad \hbox{or}\quad
\dot a_1 = -\fft{a_3}{2a_1}\,.\label{3roots}
\ee
%%%%%
The first two possibilities are associated with the first-order
equations that yield the self-dual Ricci-flat Taub-NUT metrics, whilst
the third yields the Eguchi-Hanson metric (which is also self-dual and
Ricci-flat).  In the self-dual Taub-NUT case, the $SO(3) \subset U(2)$
rotates the three hyper-K\"ahler forms as a triplet, while in the case
of the Eguchi-Hanson metrics, they are singlets under $SO(3)$.

   For future reference, we note that the equations (\ref{biaxialfo})
imply that the Weyl tensor of $ds_4^2$ satisfies the relation
%%%%%
\be 
Y^2= X^3\,,\label{xyrel}
\ee
%%%%%
where 
%%%%%
\be
X\equiv \ft1{24} C_{abcd}\, C^{abcd}\,,\qquad
Y\equiv \ft1{48} C_{abcd}\, C^{cdef}\, C_{ef}{}^{ab}\,.\label{xydef}
\ee
%%%%%

\subsection{Self-dual Taub-NUT-de Sitter metrics}

    The general biaxial Bianchi IX Einstein metrics have long been
known; these are the Taub-NUT-de Sitter solutions.  Their local form can
straightforwardly be derived by directly solving the Einstein
equations in a suitable coordinate gauge.  Writing (\ref{bianchi9}) as
%%%%%
\be
ds_4^2 = \fft{dr^2}{a_3^2} + a_1^2\, (\sigma_1^2 + \sigma_2^2) 
+a_3^2\, \sigma_3^2\,,
\ee
%%%%%
the Ricci tensor is given (in the natural orthonormal frame) by
%%%%%
\bea
R_{00}&=&-a_3\, a_3'' -{a_3'}^2 -
 \fft{2(a_1'\, a_3' \, a_3 + a_1''\, a_3^2)}{a_1}\,,\nn\\
R_{11}&=& R_{22} = -\fft{a_1''\, a_3^2}{a_1} -
  \fft{{a_1'}^2\, a_3^2}{a_1^2} -\fft{2 a_1'\, a_3'\, a_3}{a_1} 
    -\fft{a_3^2}{2a_1^4} + \fft1{a_1^2}\,,\label{ricci4}\\
R_{33}&=& -a_3\, a_3'' -{a_3'}^2 -\fft{2 a_1'\, a_3'\, a_3}{a_1} 
  +\fft{a_3^2}{2 a_1^4}\,.\nn
\eea
%%%%%
From this we see that $R_{00}-R_{33}=-a_3^2\, 
(a_1^{-4} + 4 a_1''\, a_1^{-1})/2$, and since this must vanish by the
Einstein condition, it is easy to solve for $a_1$, and hence, using the
remaining Einstein equations, for $a_3$.  Any Einstein solution to 
(\ref{ricci4}) is by definition a Taub-NUT-de Sitter metric.  Apart from
special limiting cases, the general solution has three parameters  
that we can think of as the mass $m$, the NUT charge $n$, and the
cosmological constant $\Lambda$.  This general metric is given 
by\footnote{The metric (\ref{gentnds}), parameterised by $m$, $n$ and
$\Lambda$, covers an open dense set in the modulus space of solutions
of (\ref{ricci4}).  However, for special choices of relation between the
parameters, it may be necessary to change the radial coordinate $r$
because (\ref{gentnds}) degenerates unless a limit is taken.}
%%%%%
\be
ds_4^2 = \fft{r^2-n^2}{\Delta}\, dr^2 + \fft{4n^2\, \Delta}{r^2-n^2}\, 
\sigma_3^2 + (r^2-n^2)\, (\sigma_1^2 + \sigma_2^2)\,,\label{gentnds}
\ee
%%%%%
where 
%%%%%
\be
\Delta \equiv r^2-2m\, r + n^2 + \Lambda\, (n^4 + 2 n^2 \, r^2 -
\ft13 r^4)\,.
\ee
%%%%%

   The metric (\ref{gentnds}) has a self-dual or anti-self-dual Weyl
tensor if \cite{gibpopcp2}
%%%%%
\be
m=\pm n\, (1+\ft43 \Lambda\, n^2)\,,\label{sdweyltn}
\ee
%%%%%
in which case we find 
%%%%%
\be
\Delta= (r\mp n)^2\, (1 - \ft13\Lambda(r\mp n)(r \pm n))\,.\label{sdspec}
\ee
%%%%%

     Making the specific choice of the upper sign, we obtain
the self-dual Taub-NUT-de Sitter metric
%%%%%
\be
ds_4^2 = \fft{dr^2}{F} + 4n^2\, F\, \sigma_3^2 + (r^2-n^2)\, (
\sigma_1^2 + \sigma_2^2)\,,\label{sdtndsmetric}
\ee
%%%%%
where
%%%%%
\bea
F &=& \Big(\fft{r-n}{r+n}\Big)\, (1-\ft13\Lambda(r-n)(r+3n))\,,\nn\\
&=&\fft{\Lambda}{3}\, \Big(\fft{r-n}{r+n}\Big)\,(r_+ - r)(r-r_-)\,,\qquad
r_\pm\equiv -n \pm \sqrt{4n^2+ \fft{3}{\Lambda}}\,.\label{fsdtnds}
\eea
%%%%%
The Weyl tensor is given by
%%%%%
\bea
&&C_{0101}=C_{2323}=-C_{0123}=-\fft{n\, (1+\ft13\Lambda\, n^2)}{(r+n)^3}
\,,\nn\\
&& C_{0303}=C_{1212}= -C_{0312}= 
\fft{2n\, (1+\ft13\Lambda\, n^2)}{(r+n)^3}
\,.
\label{tndsweyl}
\eea
%%%%%

   It can easily be verified that this metric satisfies the
first-order equations (\ref{foeq}).  Note that because it is biaxial,
and thus satisfies our reduced first-order system (\ref{biaxialfo}),
it follows that the Weyl tensor of the self-dual Taub-NUT-de Sitter
metrics obeys the relation (\ref{xyrel}).  It is evident that if we
send $\Lambda$ to zero in (\ref{sdtndsmetric}), we obtain the
self-dual Taub-NUT metric first written down as a Euclidean-signature
metric in \cite{hawking}:
%%%%%
\be
ds_4^2 = \Big(\fft{r+n}{r-n}\Big)\, dr^2 + 4n^2\, \Big(\fft{r-n}{r+n}\Big)
\, \sigma_3^2 + (r^2-n^2)(\sigma_1^2+\sigma_2^2)\,.\label{sdtnmetric}
\ee
%%%%%
We saw, however, that the first-order
equations (\ref{biaxialfo}) have three branches, and in the limit
where $\Lambda$ goes to zero two of these should lead to the self-dual
Taub-NUT metric, whilst the third should lead instead to the
Eguchi-Hanson metric.  As noted above, the metric form (\ref{gentnds})
with parameters $m$, $n$ and $\Lambda$, and radial coordinate $r$,
does not necessarily cover all regions of the modulus space, and in
the present case the existence of three branches suggests that there
should exist a different parameterisation of biaxial self-dual
Einstein metrics whose limiting form when $\Lambda$ goes to zero is
the Eguchi-Hanson metric.

   The required metrics cannot be the ones found
in \cite{gibpopcp2}, which are referred to as the 
Eguchi-Hanson-de Sitter metrics,
%%%%%
\be
ds_4^2=\fft{dr^2}{F} + \ft14 r^2\, F\, \sigma_3^2 + \ft14 r^2\, (
\sigma_1^2 + \sigma_2^2)\,,\label{ehds}
\ee
%%%%%
where $F=1-\ell^4\, r^{-4} - \ft16\Lambda\, r^2$, because these
metrics have neither self-dual nor anti-self-dual Weyl tensor, when
$\Lambda$ and $\ell$ are both non-zero, and thus they do not satisfy
(\ref{foeq}).  They are in fact Einstein-K\"ahler, and the Weyl tensor
has a definite duality only if $\ell=0$ (giving the Fubini-Study
metric on $\CP^2$ if $\Lambda>0$, and the Bergmann metric on the open
ball in $\C^2$ if $\Lambda<0$), or if $\Lambda=0$, in which case the
Weyl tensor has the opposite duality and the metric is
Eguchi-Hanson.\footnote{We shall discuss Bianchi IX Einstein-K\"ahler
metrics briefly in appendix \ref{ekmetsec}.}  In order to find the
``missing'' metrics, which we shall distinguish from (\ref{ehds}) by
giving them the name ``self-dual Eguchi-Hanson-de Sitter,'' it is
helpful to study the first-order equations (\ref{biaxialfo}) in
greater detail.  This forms the topic of the next subsection.

\subsection{Biaxial first-order equations, and self-dual Eguchi-Hanson-de 
Sitter}\label{sdehdssec}

    To proceed with studying the biaxial first-order equations
(\ref{biaxialfo}), we define
%%%%%
\be
u\equiv \dot a_1 + \fft{a_3}{2a_1}\,.\label{udef}
\ee
%%%%%
The cubic equation for $\dot a_1$ now becomes
%%%%%
\be
3 u^3 + (\Lambda\, a_1^2 - 3)\, u - \Lambda\, a_1\, a_3=0\,.\label{ueq}
\ee
%%%%%
One approach is to follow Cardano's procedure for solving the cubic
equation, but other than establishing the principle that there will be
two roots whose $\Lambda\longrightarrow 0$ limit yields the self-dual
Taub-NUT first-order equation $u=\pm1$, with the third yielding the
Eguchi-Hanson first-order equation $u=0$ (see (\ref{3roots})), the
direct solution of the cubic equation is not very enlightening.

   A more profitable route is to view (\ref{ueq}) as an equation
expressing $a_3$ in terms of $u$, 
%%%%%
\be
a_3 = -\fft{u\,(u^2 + \lambda\, a_1^2 -1)}{\lambda\, a_1}\,.\label{ua3}
\ee
%%%%%
Note that we are defining 
%%%%%
\be
\lambda\equiv \ft13\Lambda
\ee
%%%%%
for convenience.  In view of (\ref{ua3}), we can now choose to regard
$(a_1,u)$ as our two metric functions, rather than $(a_1,a_3)$.  From
the first-order equations (\ref{biaxialfo}) we can now deduce that
$a_1$ and $u$ satisfy the first-order equations
%%%%%
\be
\dot u = -\lambda\, a_1\,,\qquad \dot a_1 = \fft{u\, (u^2 + 3\lambda\, 
    a_1^2 -1)}{2\lambda\, a_1^2}\,.\label{ua1fo}
\ee
%%%%%

   In order to find the solution that gives rise to Eguchi-Hanson in
the $\lambda\equiv \ft13 \Lambda\longrightarrow 0$ limit, is useful to
make a redefinition that casts the equations (\ref{ua1fo}) and (\ref{ua3}) 
into a form where this limit can be taken smoothly, and such that $u$ tends
to zero in the limit.  This is easily done, by letting $u=\lambda\, w$.
The first-order equations (\ref{ua1fo}) become
%%%%%
\be
\dot w = -a_1\,,\qquad \dot a_1 = \fft{w\, (\lambda^2\, w^2 + 
   3\lambda\, a_1^2-1)}{2a_1^2}\,,\label{ehbranch}
\ee
%%%%%
and (\ref{ua3}) gives
%%%%%
\be
a_3 = - \fft{w\, (\lambda^2 \, w + 3\lambda\, a_1^2 -1)}{a_1}\,.
\ee
%%%%%
It follows that the solution to (\ref{ehbranch}) for general
non-vanishing $\lambda$ will give the required self-dual
Eguchi-Hanson-de Sitter metrics.  By defining a new radial variable
$x$ such that $dx=-a_1\, dt$, the equation for $w$ can be solved
to give $w=x$, and hence the solution for $a_1$ can be found.  After
a further simple coordinate redefinition, the solution can be
expressed as
%%%%%
\be
ds_4^2 = \fft{d\rho^2}{U\, V} + \ft14 \rho^2\, \fft{V}{1-2\mu\, \ell^2}\, 
  (\sigma_1^2 + \sigma_2^2) + \ft14 \rho^2\, \Big(\fft{1-\mu\, \ell^2}{
1-2\mu\, \ell^2}\Big)^2 \, U\, V \, \sigma_3^2\,,\label{sdehdsmet}
\ee
%%%%%
where
%%%%%
\be
U\equiv 1-\fft{\ell^4}{\rho^4}\,,\qquad V\equiv 1-\mu\, (\ell^2+\rho^2)\,.
\ee
%%%%%
The metric is Einstein, with cosmological constant $\Lambda=12\mu$, 
and its Weyl tensor is anti-self-dual.  In fact, we find that the
tangent-frame components of the Weyl tensor are given by
%%%%%
\bea
&&C_{0101}=C_{2323}=-C_{0123}= -\fft{2\ell^4\,(1-\mu\, \ell^2)}{\rho^6}
\,,\nn\\
&& C_{0303}=C_{1212}= -C_{0312}= \fft{4\ell^4\, (1-\mu\, \ell^2)}{\rho^6}\,.
\label{ehdsweyl}
\eea
%%%%%
 
  Since (\ref{sdehdsmet}) is Einstein and of biaxial Bianchi IX type,
it must be contained within the general Taub-NUT-de Sitter class of
solutions (\ref{gentnds}).  Furthermore, since its Weyl tensor is
anti-self-dual, it can be expected to lie within the subclass of
(\ref{gentnds}) that satisfy (\ref{sdweyltn}).  After simple algebra
we find that there is indeed a transformation that maps
(\ref{gentnds}) with the anti-self-dual specialisation given by
(\ref{sdweyltn}) into (\ref{sdehdsmet}), namely 
%%%%%
\be
r^2=-\fft{(12-\Lambda\, \ell^2 -2\Lambda\, \rho^2)^2}{32\Lambda\, 
    (6-\Lambda\, \ell^2)}\,,\qquad
n^2 = -\fft{(12-\Lambda\, \ell^2)^2}{32\Lambda\, (6-\Lambda\, \ell^2)}\,.
\label{tn2eh}
\ee
%%%%%
Substituting these redefinitions into (\ref{gentnds}) with (\ref{sdspec}),
we recover (\ref{ehbranch}).  It should be noted that when $\Lambda\, 
\ell^2< 6$, the self-dual Eguchi-Hanson-de Sitter metric corresponds to 
a section of the self-dual Taub-NUT-de Sitter metric in which the NUT 
parameter and radial coordinate are imaginary.  Thus from the point 
of view of the real geometry, the self-dual Taub-NUT-de Sitter and 
self-dual Eguchi-Hanson-de Sitter metrics should be viewed as inequivalent.

    In order to clarify the relations between the self-dual
Taub-NUT-de Sitter and self-dual Eguchi-Hanson-de Sitter metrics, and
more generally to investigate the full solution space of the self-dual
biaxial metrics, it is useful to study the phase-plane for the
first-order system (\ref{ua1fo}). Before doing so, we shall close this
subsection by showing where two well-known self-dual Einstein metrics
that are contained within the biaxial Bianchi IX class fit in, namely 
$S^4$ and $\CP^2$.

   Setting $\ell=0$ in (\ref{sdehdsmet}) gives $S^4$, as can be seen
by changing to the radial coordinate $t$ defined by $\sqrt\mu\,
\rho=\sin\ft12 t$.  This gives
%%%%%
\be
ds_4^2 =\fft{3}{\Lambda}\, (dt^2 + \ft14 \sin^2 t\, \sigma_i^2)\,.
\label{s4met}
\ee
%%%%%
From (\ref{tn2eh}), this corresponds to
$n^2=-3/(4\Lambda)$ in the self-dual Taub-NUT-de Sitter
parameterisation.  

   Another special case of (\ref{sdehdsmet}), which arises when
$\Lambda\, \ell^2=12$, also gives rise to $S^4$.  This is a singular
limit, for which we must first rescale the Euler angle $\psi$ that
appears in $\sigma_3=d\psi + \cos\theta\, d\phi$ according to
$\psi=(1-\ft1{12}\Lambda\, \ell^2)^{-1}\, \tau$.  Substituting into
(\ref{sdehdsmet}), and then sending $\ell^2 \longrightarrow
12/\Lambda$, we obtain
%%%%%
\be
ds_4^2 = \fft{3}{\Lambda}\, [d\chi^2 + \sin^2\chi \, (\sigma_1^2 
+\sigma_2^2) + \cos^2\chi\, d\tau^2 ]\,,\label{s4again}
\ee
%%%%%
where we have also set $\Lambda\, \rho^2 = 12\sin\chi$.  We can
recognise (\ref{s4again}) as the metric on $S^4$, written as a
foliation by $S^2\times S^1$ surfaces.  The fact that
(\ref{sdehdsmet}) describes $S^4$ both for $\ell^2=0$ and $\ell^2=
12/\Lambda$ is not unexpected in view of the expressions (\ref{ehdsweyl}),
since the Weyl tensor can be seen to vanish for these two values of
$\ell^2$.  Note that from (\ref{tn2eh}) the value of NUT parameter in
the self-dual Taub-NUT-de Sitter parameterisation corresponding to the
$S^4$ limit with $\Lambda\, \ell^2=12$ is $n=0$.

   A further special case of (\ref{sdehdsmet}) is when $\Lambda\,
\ell^2=6$.  This gives $\CP^2$.  One must first define a new radial
coordinate, for example by setting $\rho^2= (2\mu)^{-1}\, [1+(1-2\mu\,
\ell^2)\, \cos 2\chi]$, before taking the limit.  We then obtain the
metric
%%%%%
\be
ds_4^2 = \fft{6}{\Lambda}\, [d\chi^2 + \ft14 \sin^2\chi\, (\sigma_1^2 
+\sigma_2^2) + \ft14 \sin^2\chi\,\cos^2\chi\, \sigma_3^2]\,,\label{cp2}
\ee
%%%%%
which can be recognised as the Fubini-Study metric on $\CP^2$
\cite{gibpopcp2}.  From (\ref{tn2eh}), it corresponds, in the 
self-dual Taub-NUT-de Sitter parameterisation, to sending the NUT
parameter $n$ to infinity.  Note that with the conventions of this
paper, the Weyl tensor is anti-self-dual, as is the (covariantly
constant) K\"ahler form $J=e^0\wedge e^3 -e^1\wedge e^2$.  Of course
none of the self-dual quaternionic K\"ahler forms $J^i$ is covariantly
constant, since the right-handed $SU(2)$ part of the spin connection
is non-vanishing.

\subsection{Phase-plane analysis for the biaxial system}\label{phpl1}

    As we have seen above, finding a uniform parameterisation of the
space of solutions, even in the biaxial case, is non-trivial.  The approach
taken in this section will be to classify all the possible orbits in
the phase space of the first-order equations (\ref{ua1fo}).  We shall 
find that not all solutions can be parameterised by giving real and finite
values of $\ell$, or $n$.

    We begin by making the definition $v= 3 a_1$, choosing the scale
size $\lambda=3$ for convenience, and sending $t\longrightarrow -t$
for inconvenience.  The first-order equations (\ref{ua1fo}) become
%%%%%
\be
\dot u=v\,,\qquad \dot v = \fft{9 u\, (1-u^2-v^2)}{2 v^2}\,,\label{uvfo}
\ee
%%%%%
and so the solutions can be represented as flows in the
$(u,v)$ plane.  We can divide the two equations to get
%%%%%
\be
\fft{dv}{du} =  \fft{9 u\, (1-u^2-v^2)}{2 v^3}\,.\label{dvdu}
\ee
%%%%%

   In general, equation (\ref{dvdu}) can be integrated to
give the flows for any biaxial self-dual solution.  The constant
of integration is related to the NUT parameter $n$, or, equivalently,
the scale parameter $\ell$ in the self-dual Eguchi-Hanson-de Sitter 
formulation.  In terms of $n$, the integral of (\ref{dvdu}) is given by
%%%%%
\be
(u^2+\ft13 v^2 -1)^2 = 12n^2\, (u^2+\ft23 v^2 -1)\,.
\label{uvsol}
\ee
%%%%%
Since this is symmetrical under reflections in the $u$ and $v$ axes,
it suffices to consider flows within the positive quadrant.  

   It follows from (\ref{dvdu}) that flow lines inside the unit circle
have positive gradient, whilst those outside the unit circle have
negative gradient.  The $v$ axis corresponds to $a_3=0$, signifying an
endpoint of the metric at which the 3-dimensional orbits degenerate to
an $S^2$ bolt.  The $u$ axis, on the other hand, corresponds to
$a_1=0$, and the metric will be singular here unless it happens that
$u=\pm1 $ or $u=0$, in which case the orbits degenerate to a point,
implying a NUT endpoint in the metric.  By a theorem of Hitchin's, the
only complete and non-singular metrics with positive $\Lambda$ are
$S^4$ and $\CP^2$.

   It is straightforward to establish that the $\CP^2$ solution
(\ref{cp2}) corresponds to the ellipse $u^2 + \ft23 v^2 =1$.  The flow
starts on the $v$ axis at $v^2=\ft32$ at a bolt, and runs along the
ellipse to a NUT on the $u$ axis at $u=1$.  Since we have chosen the 
normalisation $\mu=\ft1{12}\, \Lambda=\ft14\lambda =\ft34$ in this subsection,
it follows that this occurs for $\ell^2 =\ft23$. 

   The $S^4$ solution (\ref{s4met}) with $\ell^2=0$ corresponds to the
ellipse $(u-\ft12)^2 +\ft13 v^2 =\ft14$.  This runs from the NUT at
$u=0$, $v=0$ to the NUT at $u=1$, $v=0$.  The other $S^4$ solution
(\ref{s4again}), with $\ell^2= 12/\Lambda=4/3$, corresponds to the ellipse
$u^2+\ft13 v^2 =1$.  Although this appears to be singular since, from
(\ref{ua3}), we have $a_3=0$, we saw that to obtain (\ref{s4again}) it
was necessary to rescale the $\psi$ coordinate and this has the effect of
compensating for the vanishing of $a_3$.

    The phase-plane plot, with the various ellipses and unit circle
mentioned above displayed, is given in Figure 1.  

\bigskip

\begin{figure}[ht]
\epsfxsize=3.4truein
\leavevmode\centering
\epsfbox{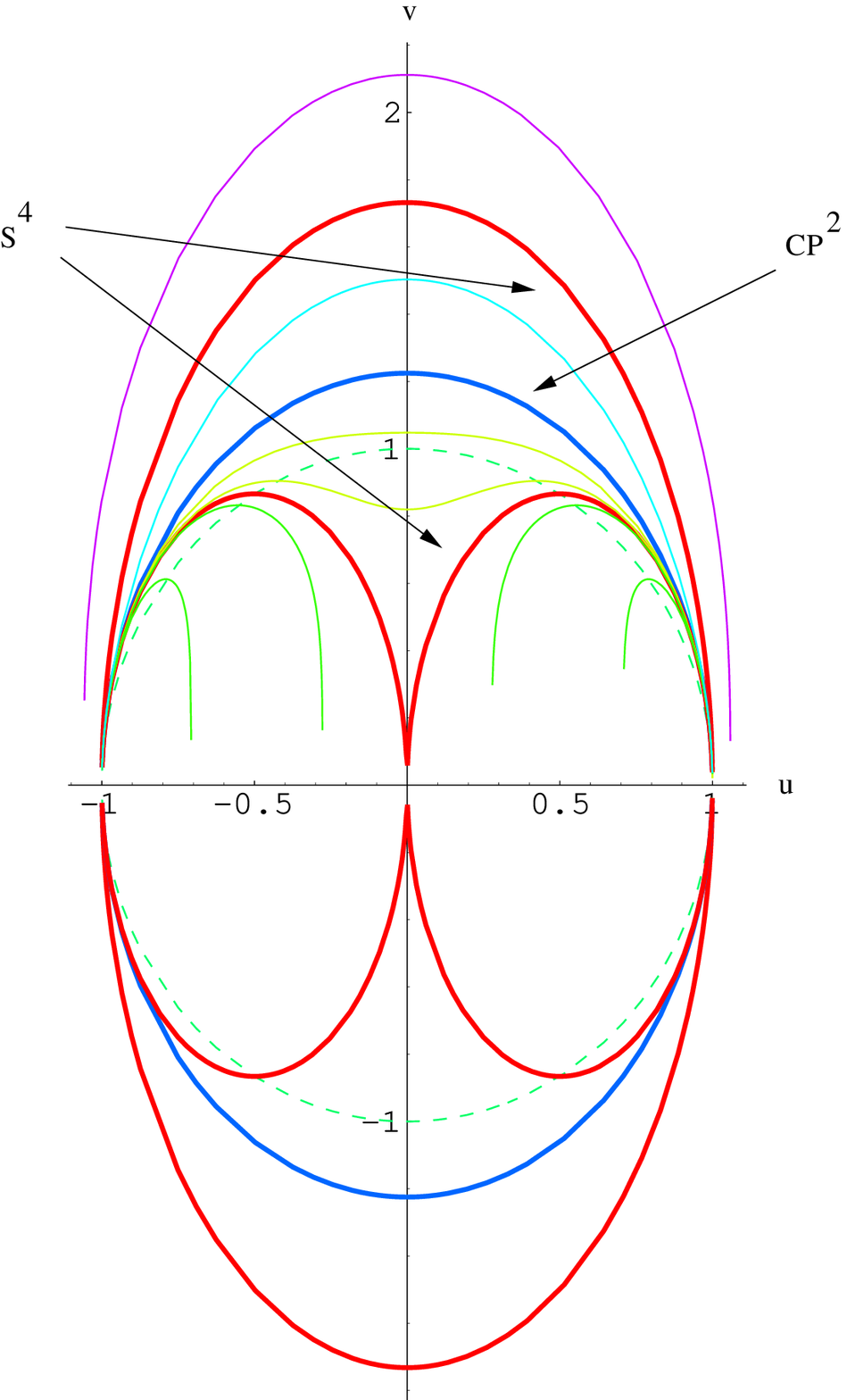}
\caption{The phase plane for the first-order system of equation
(\ref{uvfo}).  The heavy blue ellipse corresponds to the $\CP^2$ flow,
and the three heavy red ellipses to $S^4$ flows.  The dashed green
circle is $u^2+v^2=1$; all solutions that cross this do so
horizontally.  To label distinct metrics it is sufficient to consider
flows lying within the positive quadrant.  We label qualitatively
similar flows by A, B, C and D, which indicate the regions they occupy
and their initial points.  Thus the regions A, B and C indicate
starting-points for solutions on the $v$ axis.  Region A ranges from
$v=0$ to the intersection of the $\CP^2$ ellipse with the $v$
axis. Region B ranges from this intersection to the intersection of
the outer ellipse (the $S^4$ solution (\ref{s4again})) with the $v$
axis.  Region C ranges from this point to $v=+\infty$. Region D
denotes (singular) starting points on the $u$ axis for solutions, in
the range $0<u<1$. For clarity we have plotted complete ellipses for
the $S^4$ an $\CP^2$ special cases, but only flows in the upper
half-plane for the other representative examples.}
\end{figure}
%\clearpage

   From (\ref{uvsol}), we see that solutions starting from a bolt
on the $v$ axis are specified at $u=0$ by
%%%%%
\bea
\hbox{\bf Region A}:&& -\infty <n^2 <-\ft1{12}:
\qquad \qquad \ft32 > v^2 >0\,,\nn\\
\hbox{\bf Region B}:&& -\infty < n < 0:\qquad\qquad
   \ft32 < v^2 < 3 \,,\nn\\
\hbox{\bf Region C}:&& 0 < n < \infty :\qquad\qquad
    3 < v^2 < \infty \,.\nn
\eea
%%%%%
We also have solutions starting from the singular curvature singularity
along the $u$ axis, specified at $v=0$ by:
%%%%%
\be
\hbox{\bf Region D}:\qquad -\ft1{12} < n^2 <0:\qquad\qquad
0<u<1\,.
\ee
%%%%%
In terms of the parameter $\ell^2$ of the self-dual Eguchi-Hanson-de Sitter
formulation of the metrics, we see from (\ref{tn2eh}) that Region A,
where $n$ is imaginary, is covered by real values of $\ell$, and so
the self-dual Eguchi-Hanson-de Sitter form of the metrics is better adapted
to describing this region of the phase plane.  On the other hand, in 
Region D, where $n$ is again imaginary, $\ell$ is complex, and so neither
the self-dual Taub-NUT-de Sitter nor self-dual Eguchi-Hanson-de Sitter
formulation is well adapted to describing this region of the phase plane.
It is straightforward to find an adapted parameterisation where the 
analogue of the NUT parameter, and the radial coordinate, is real in
Region D, but since the metrics there have power-law curvature singularities
there is not much value in writing them down.

    It is instructive to express the Weyl tensor for the biaxial self-dual
metrics in terms of $u$ and $v$.  We find that it is given by
%%%%%
\be
C_{0101}=-C_{0123}= C_{2323}=-2C_{0303}=2C_{0312}= -2C_{1212}=f\,,
\label{weyltensor}
\ee
%%%%%
where
%%%%%
\be
f=-\ft{81}{2}\, v^{-6}\, (u^2+ \ft13 v^2 -1)\,
\Big((u-\ft12)^2 + \ft13 v^2 -\ft14\Big)\,
\Big((u+\ft12)^2 + \ft13 v^2 -\ft14\Big)\,.
\ee
%%%%%
As expected, this vanishes only on the $S^4$ ellipses, and it diverges
everywhere on the $u$ axis except at the points $u=0,\pm1$, provided
they are approached along the $S^4$ flows.

\subsection{Global structure of the biaxial solutions}

    As we have already remarked, a theorem of Hitchin's implies
that when the cosmological constant $\Lambda$ is positive, 
only the $S^4$ and $\CP^2$ self-dual Einstein metrics can be non-singular.
In particular, therefore, this means that the self-dual Taub-NUT-de 
Sitter and self-dual Eguchi-Hanson-de Sitter metrics will be 
singular except for the special values of $n$ or $\ell^2$ for which
they reduce to $S^4$ or $\CP^2$.

    We shall analyse the self-dual Taub-NUT-de Sitter metrics first,
described by (\ref{sdtndsmetric}) and (\ref{fsdtnds}).  The
coordinate $r$ is taken to lie in the interval $n\le r\le r_+$. 
For convenience, we shall again set $\Lambda=9$ here.
Near $r=n$,  letting $r-n=\rho^2$, the metric becomes
%%%%%
\be
ds_4^2\sim 8n\, [d\rho^2 + \ft14\rho^2\, (\sigma_1^2 + \sigma_2^2 + 
\sigma_3^2)]\,,
\ee
%%%%%
which describes a NUT.  The metric smoothly approaches the origin of
$\R^4$, provided that the Euler angle $\psi$ appearing in
$\sigma_3=d\psi + \cos\theta\, d\phi$ has its canonical period $4\pi$.

    Near $r=r_+$, by letting $r_+-r=\rho^2$ we see that the metric becomes
%%%%%
\be
ds_4^2 \sim \fft{2}{(\sqrt3 \sqrt{12n^2+1} -6n)}\, 
( d\rho^2 + 4n^2\, (\sqrt3 \sqrt{12n^2+1} -6n)^2\, \rho^2\, \sigma_3^2) + 
(r_+^2-n^2)\, (\sigma_1^2+\sigma_2^2)\,.
\ee
%%%%%
This approaches $\R^2\times S^2$ locally, but in general there will
be a conical singularity.  If $\psi$ has period $\Delta\psi=4\pi/N$, 
then regularity at $r=r_+$ is achieved if
%%%%%
\be
N=4n(-6n+\sqrt3\sqrt{12n^2+1})\,.\label{Nnrel}
\ee
%%%%%
 
    Regularity at $r=n$ required $N=1$.  This is compatible with 
(\ref{Nnrel}) if $n=\infty$, which is the limit where the
self-dual Taub-NUT-de Sitter metric becomes $\CP^2$ \cite{gibpopcp2}.
(Another case where the singularity can be avoided is by taking a
limit where $n^2\longrightarrow -3/(4\Lambda)=-\ft1{12}$, in which case 
one must first rescale coordinates in the metric.  This case is $S^4$.)  For
all other values of $n$, there will be a deficit angle at the origin,
and a hence a conical singularity.  

  The $G_2$ metrics (\ref{g2met}) obtained by taking $ds_4^2$ to be
self-dual Taub-NUT-de Sitter were discussed recently in
\cite{behrndt}.  They have cohomogeneity 2, since there are two
``radial'' coordinates $\rho$ and $t$.  The conical singularities in
the Taub-NUT-de Sitter metrics imply, of course, that the
corresponding $G_2$ metrics will have conical singularities too.

   A further class of geometries within the biaxial Bianchi IX class is
obtained by considering instead the self-dual Eguchi-Hanson-de Sitter
form of the metrics, given by (\ref{sdehdsmet}).  If $2\mu\, \ell^2<1$,
meaning that $\Lambda\, \ell^2<6$, the radial coordinate $\rho$ can 
be chosen to lie in the interval $\ell\le \rho\le \rho_0$, where
$\rho_0^2 = 1/\mu -\ell^2$.  Near $\rho=\ell$, setting $\rho=\ell+x^2$,
we find
%%%%%
\be
ds_4^2 \sim \fft{\ell}{1-2\mu\, \ell^2}\, [dx^2 + (1-\mu\, \ell^2)^2\, 
   x^2\, \sigma_3^2]  + \ft14 \ell^2\, (\sigma_1^2 + \sigma_2^2)\,,
\ee
%%%%%
whilst near $\rho=\rho_0$, we have, setting $\rho=\rho_0-x^2$,
%%%%%
\be
ds_4^2 \sim \fft{1-\mu\, \ell^2}{1-2\mu\, \ell^2}\, (dx^2 + \ft14 x^2\, 
\sigma_i^2)\,.
\ee
%%%%%
Thus regularity at the NUT at $\rho=\rho_0$ requires that $\psi$ have
period $4\pi$, which implies that there is a conical singularity on the
bolt at $\rho=\ell$.

\subsection{Phase plane and global structure for negative $\Lambda$}

   The phase-plane analysis of section \ref{phpl1} can be repeated for
the case where the cosmological constant $\Lambda$ is taken to be
negative.  Starting from (\ref{ua1fo}) and (\ref{ua3}), and fixing the
scale by choosing $\lambda\equiv \ft13\Lambda=-3$, we now have
%%%%%
\bea
&&\dot u = v\,,\qquad \dot v = \fft{9u\, (1-u^2+v^2)}{2v^2}\,,\nn\\
&&\fft{dv}{du} =  \fft{9u\, (1-u^2+v^2)}{2v^3}\,,\nn\\
&&a_3 = -\fft{u}{v}\, (1-u^2 + \ft13 v^2)\,.\label{neglam}
\eea
%%%%%

   The flow can be integrated, giving
%%%%%
\be
(u^2-\ft13 v^2 -1)^2 + 12 n^2 \, (u^2-\ft23 v^2 -1)=0\,.
\ee
%%%%%
As in the case when $\Lambda>0$, this is symmetric under reflections in
the $u$ and $v$ axes.
The hyperbola $u^2-\ft23 v^2=1$, which arises when $n=\infty$,
corresponds to the Bergmann metric on the open ball in $\bC^2$ (\ie
the Fubini-Study metric with negative $\Lambda$, which is the coset
$SU(2,1)/U(2)$).  The hyperbolic 4-space $H^4$ arises if $n=0$, giving
the hyperbola $u^2-\ft13 v^2=1$.  It also arises if $n^2=\ft1{12}$, 
giving the hyperbolae $(u\pm \ft12)^2 -\ft13 v^2 =\ft14$.

  The Weyl tensor is given by (\ref{weyltensor}), where $f$ is now given
by
%%%%%
\be
f=\ft{81}{2}\, v^{-6}\, (u^2- \ft13 v^2 -1)\,
\Big((u-\ft12)^2 - \ft13 v^2 -\ft14\Big)\,
\Big((u+\ft12)^2 - \ft13 v^2 -\ft14\Big)\,.
\ee
%%%%%
The Weyl tensor therefore vanishes on the $H^4$ hyperbolae, and has a
power-law divergence at all points on the $u$ axis except if one
approaches $u=0,\pm1$ along the $H^4$ flows.

   Writing the metric in the self-dual Eguchi-Hanson-de Sitter form
(\ref{sdehdsmet}), where now $\mu\equiv \ft1{12}\Lambda$ is taken to
be negative, say $-\mu\equiv \nu >0$, we see that the radial
variable can be taken in the range $\rho\ge \ell$.  Near $\rho=\ell$
we set $\rho=\ell + x^2$, giving
%%%%%
\be
ds_4^2 \sim \fft{\ell}{1+2\nu\, \ell^2}\, [dx^2 + (1+\nu\, \ell^2)^2\,
   x^2\, \sigma_3^2] + \ft14\ell^2\, (\sigma_1^2 + \sigma_2^2)\,.
\ee
%%%%%
Thus we have a regular $S^2$ bolt, provided that the period $\Delta\psi$
of $\psi$ is chosen to be 
%%%%%
\be
\Delta\psi = \fft{2\pi}{1+ \nu\, \ell^2}\,.
\ee
%%%%%
Provided that $\ell$ is such that this period is $4\/N$, for $N$ an
integer, we shall have a regular metric, with $S^3/Z_N$ orbits.

   Now consider instead writing the metric in the self-dual
Taub-NUT-de Sitter form (\ref{sdtndsmetric}).  Taking $\Lambda=-9$ for
simplicity, the roots $r_\pm$ are given by $r_\pm =-n
\pm\ft1{\sqrt3}\, \sqrt{12n^2-1}$.  Assuming $n^2>\ft1{12}$, this means
that the roots $r_\pm$ are both less than $n$ (assumed positive), and
so we can take $r\ge n$.  Near $r=n$ we set $r=n+x^2$, finding
%%%%%
\be
ds_4^2 \sim 8n\, (dx^2 + \ft14 x^2\, \sigma_i^2)\,.
\ee
%%%%%
Thus $r=n$ is a regular NUT, provided $\psi$ has period $4\pi$.

   The regular solutions with a bolt, which we described in the 
self-dual Eguchi-Hanson-de Sitter form (\ref{sdehdsmet}) above, can also
be expressed in the self-dual Taub-NUT-de Sitter form.  They correspond
to running the radial coordinate $r$ from $r=r_-$ to $r=-\infty$ (note
that $r_-<-n$, so the curvature singularity at $r=-n$ is avoided).  

   All the other solutions represented in Figure 2 have flows that intersect
the $u$ axis at points other than $u=0$ or $\pm1$, and thus they have
power-law curvature singularities.

\bigskip

\begin{figure}[ht]
\epsfxsize=4truein
\leavevmode\centering
\epsfbox{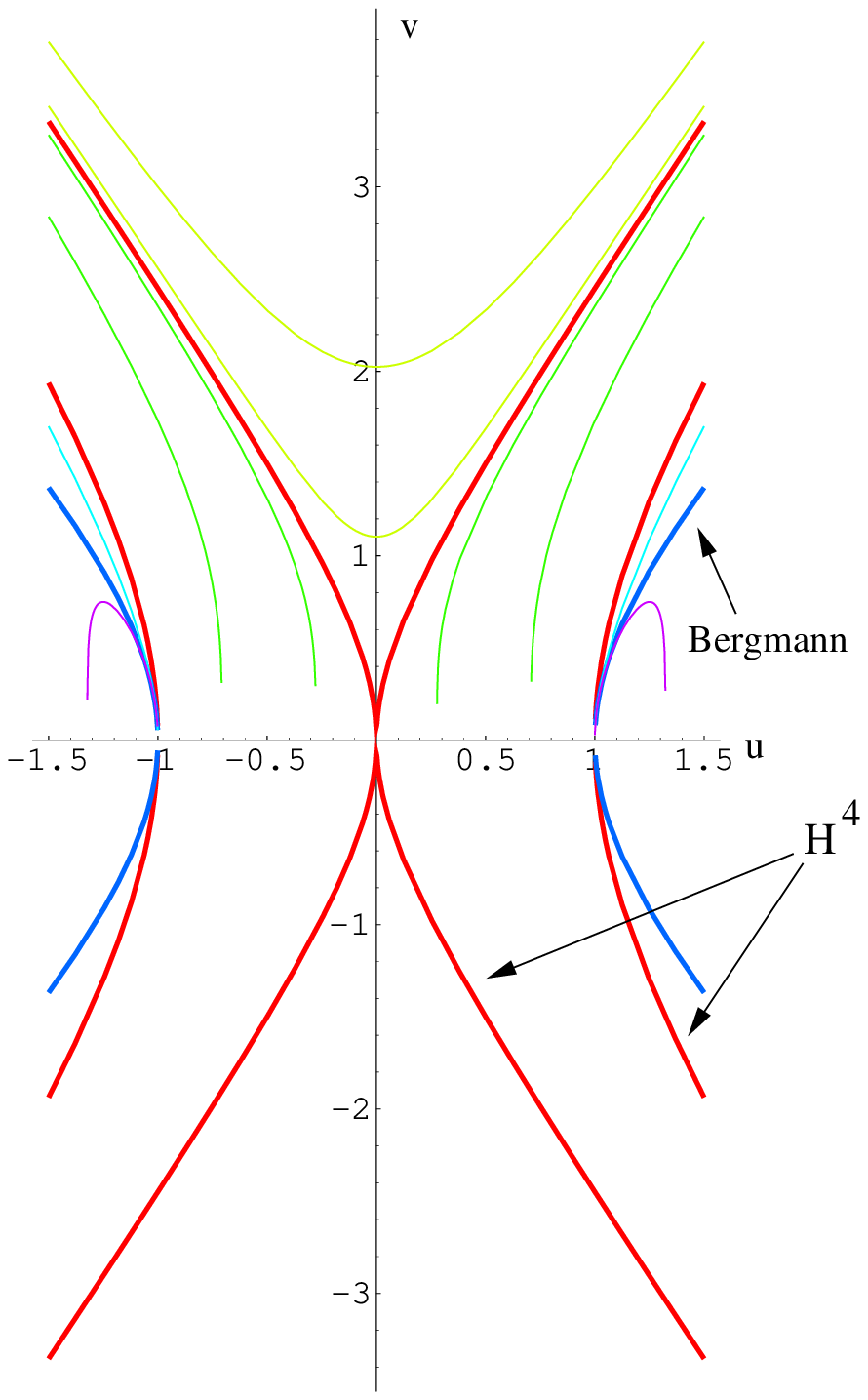}
\caption{The phase plane for the first-order system of equation
(\ref{neglam}). The heavy blue hyperbola corresponds to the Bergmann flow, and
the three heavy red hyperbolae to $H^4$ flows.  The discussion of 
other flows is analogous to that for $\Lambda>0$; some representative 
examples are depicted.}

\end{figure}
%\clearpage

\subsection{Superpotential for the biaxial system}\label{superpotsec}

     Although the $D=4$ self-dual Einstein spaces do not themselves
have special holonomy, the existence of the first-order system implies
that it might be possible to derive it from a superpotential.  To
obtain such a superpotential, we first notice that the Hamiltonian of
the cohomogeneity one Einstein space is given by $H=T + U$, where
%%%%%
\bea
T&=& \fft{2a_1'\, a_2'}{a_1\, a_2} + \fft{2a_1'\, a_3'}{a_1\, a_3} +
\fft{2a_2'\, a_3'}{a_2\, a_3}\,,\nn\\ 
U&=&\ft12(a_1^4 + a_2^4 + a_3^4
- 2 a_1^2\, a_2^2 - 2 a_1^2\, a_3^2 -2a_2^2\, a_3^2 + 12\lambda\,
a_1^2\, a_2^2\, a^2_3)\,, \label{kinpot}
\eea
%%%%%
and a prime denotes a derivative with respect to $\eta$ defined by
$dt=a_1\, a_2\, a_3\, d\eta$. 

   Here, we shall consider the biaxial system with
$a_2=a_1$, and use the $u$ and $a_1$ variables defined in section
\ref{sdehdssec}.  We can write $T=\ft12 g_{ij}\, (d\alpha^i/d\eta)\, 
(d\alpha^j/d\eta)$, with
$\alpha_i=(a_1, u)$, which implies that $g_{ij}$ is given by
%%%%
\bea
g_{ij}&=&\pmatrix{\fft{4x}{a_1^2 y} & \fft{4 z}{a_1\, u\, y} \cr
              \fft{4z}{a_1\, u\, y} & 0\cr}\,,\nn\\
x&\equiv &1 + 3\lambda\, a_1^2 - u^2\,,\quad
y\equiv -1 +\lambda\, a_1^2 + u^2\,,\quad
z\equiv -1 + \lambda\, a_1^2 + 3u^2\,.
\eea
%%%%
We find that the potential $U$ can then expressed as
$U=-\ft12 g^{ij}\, (\del W/\del \alpha_i)\, (\del W/\del\alpha_j)$,
with the superpotential $W$ given by
%%%%
\be
W= -\fft{u^2\, (u^2-1)^2}{\lambda^2\, a_1^2} 
 -\fft{ u^2\, (5u^2-4)}{\lambda} -
a_1^2\, (3u^2 +2) + \lambda\, a_1^4\,.\label{biaxsup}
\ee
%%%%%
It is straightforward to derive the first order equations from this
superpotential.

\section{Triaxial anti-self-dual Bianchi IX metrics}

   In this section we discuss the full triaxial system of equations, 
which are considerably more complicated than the biaxial case.  

\subsection{Phase-plane and superpotential for triaxial system}

   We begin with an outline of a phase-plane analysis for the triaxial system,
using methods similar to those that we used for the biaxial case.

   Starting from the first-order equation for $\dot a_1$ obtained in
(\ref{a1eq}), it is natural to define the auxiliary variable $w$, 
by
%%%%%
\be
w\equiv \dot a_1 - \fft{a_1^2-a_2^2-a_3^2}{2a_2\, a_3}\,.
\label{wtriaxialdef}
\ee
%%%%%
In terms of $w$, equation (\ref{a1eq}) becomes
%%%%%
\be
(w^2-1)^2 \, (w-\lambda\, a_2\, a_3) - \lambda^2\, a_1^2\, (a_2\, w+ a_3)
(a_3\, w+a_2) =0\,.\label{quinticw}
\ee
%%%%%
The remaining first-order equations in (\ref{foeq}), namely those for
$\dot a_2$ and $\dot a_3$, then become
%%%%%
\bea
\dot a_2 &=& \fft{a_2^2 -a_1^2 -a_3^2}{2 a_1\, a_3} - \fft{\lambda\, 
a_1\, (a_2\, w+a_3)}{w^2-1}\,,\nn\\
\dot a_3 &=& \fft{a_3^2 -a_1^2 -a_2^2}{2 a_1\, a_2} - \fft{\lambda\, 
a_1\, (a_3\, w+a_2)}{w^2-1}\,.\label{a2a3eq}
\eea
%%%%%

   We can now try following the strategy of treating $(w,a_2,a_3)$ as
the independent variables, instead of $(a_1, a_2, a_3)$.  This is
similar to the strategy used in the biaxial case, although not exactly 
parallel.  Differentiating (\ref{quinticw}), using (\ref{a2a3eq}) and
(\ref{wtriaxialdef}), and then using (\ref{quinticw}) itself to substitute
for $a_1^2$, we get
%%%%%
\bea
&&(\dot w + \lambda\, a_1)(w^2-1)^2\, [a_2\, a_3\, (3w^4 +6w^2 -1)-
4(a_2^2+a_3^2)\, w^3 \nn\\
&&\qquad\qquad\qquad - \lambda\, a_2\, a_3\, ((a_2^2+a_3^2)(1+3w^2) 
      + 2 a_2\, a_3\, w\, (w^2 + 3)) ]=0\,.
\eea
%%%%%
Unless the algebraic expression contained in square brackets vanishes,
we therefore have the first-order equation
%%%%%
\be
\dot w= -\lambda\, a_1\,.
\ee
%%%%%
We should think of $a_1$ as being solved for here, using (\ref{quinticw}).
Since this would involve the use of square roots, it seems preferable 
to introduce a new radial variable $\rho$, defined by $d\rho = -a_1\, dt$.
We then have
%%%%%
\be
w' = \lambda\,.
\ee
%%%%%
The remaining first-order equations (\ref{a2a3eq}) will also involve
$a_1$ only through $a_1^2$, and so we shall have the system
\bea
%%%%%
w' &=& \lambda\,,\nn\\
a_2' &=&  -\fft{a_2^2 -a_1^2 -a_3^2}{2 a_1^2\, a_3} + \fft{\lambda\, 
(a_2\, w+a_3)}{w^2-1}\,,\nn\\
a_3' &=& -\fft{a_3^2 -a_1^2 -a_2^2}{2 a_1^2\, a_2} + \fft{\lambda\, 
(a_3\, w+a_2)}{w^2-1}\,,\label{a2a3eq2}
\eea
%%%%%
where from (\ref{quinticw}), $a_1^2$ is given by
%%%%%
\be   
a_1^2 = \fft{(w^2-1)^2\, (w-\lambda\, a_2\, a_3)}{\lambda^2\, 
(a_2\, w+a_3)(a_3\, w+a_2)}\,.
\ee
%%%%%

  Analogously to the biaxial case, we see from (\ref{quinticw})
that when $\lambda=0$ we have $w=0$ corresponding to the ``BGPP''
first-order equations, and $w=\pm1$ corresponding to the
``Atiyah-Hitchin'' first-order equations.

   The problem of solving the general triaxial first-order equations
can be reduced to a second-order equation in a single
variable. Defining $y\equiv a_2/a_3$ and $z\equiv a_2\, a_3$, we find,
after normalising so that $\lambda=1$, that $y$ satisfies the equation
%%%%%
\be
y'' - \fft{3y^2 -2 \rho\, y +1}{y\, (y^2-1)}\, {y'}^2 
-\fft{2\rho\, y^2-3\rho^2 \, y -y+2\rho}{\rho\, (\rho^2-1)\, y}\, y'
 +\fft{2(y^2-1)}{\rho\, (\rho^2-1)^2} =0\,,
\ee
%%%%%
and that $z$ is given by
%%%%%
\be
z= \fft{\rho\,(\rho^2-1)\,  y\, [(\rho^2-1)\, y' + 1-y^2]}{
(\rho^2-1)^2 \, y\, y' + \rho\, (y^4-1) +2y^2\, (y^2-1)}\,.
\ee
%%%%% 

    We find that it is possible to derive the triaxial first-order system
from a superpotential also.  We use $\a_i=(u,a_2,a_3)$ as variables, as
discussed above.  The kinetic energy $T$ given in (\ref{kinpot}) can
be straightforwardly rewritten in terms of derivatives of $(u,a_2,a_3)$, and
hence we can read off the components of the sigma-model metric $g_{ij}$
in $T=\ft12 g_{ij}\, (d\a^i/d\eta)\, (d\a^j/d\eta)$, where as before 
$dt=a_1\, a_2\, a_3\, d\eta$.  Since the expression for $g_{ij}$
is quite complicated, we shall not present it here.  Then, we find after
some calculation that the potential $U$ given in (\ref{kinpot}) can be
written in terms of a superpotential $W$ as $U=-\ft12 g^{ij}\, 
(\del W/\del\a^i)\, (\del W/\del\a^j)$, with
%%%%%
\be
W = -a_2^2 -a_3^2 -2\lambda^{-1}\, w^2 + \fft{a_2\, a_3\, (3w^2-1)}{w}
 + \fft{(\lambda^2\, a_2^2\, a_3^2 -w^2)(w^2-1)^2}{
\lambda^2\, w\, (a_2\, w+a_3)(a_3\, w+a_2)}\,.\label{triaxsup}
\ee
%%%%%
 
    It is easily verified that if one sets $a_2=a_1$, then $w$ reduces to
the function $u$ of the biaxial system, and, after using (\ref{ua3}) 
to replace $a_3$ by $u$, then $W+\lambda^{-1}$ becomes the biaxial 
superpotential given in (\ref{biaxsup}), where $W$ denotes the 
triaxial superpotential (\ref{triaxsup}) after the biaxial specialisations.

\subsection{The Tod-Hitchin first-order system}

    In this section we shall follow Tod \cite{tod} and Hitchin
\cite{hitchin,hitchin2}, who use a different approach to study the
general triaxial system (\ref{foeq}).  The metric is written as
%%%%%
\be
ds_4^2 = F\, \Big( \fft{dx^2}{x\, (1-x)} +
\fft{\sigma_1^2}{\Omega_1^2} + \fft{(1-x)\, \sigma_2^2}{\Omega_2^2}
 + \fft{x\, \sigma_3^2}{\Omega_3^2}\Big)\,.\label{ommetric}
\ee
%%%%%
Tod \cite{tod} shows that $ds_4^2$ is Einstein with anti-self-dual
Weyl tensor if the functions $\Omega_i$ satisfy
%%%%%
\be
\Omega_1'= -\fft{\Omega_2\, \Omega_3}{x\, (1-x)}\,,\quad
\Omega_2' = -\fft{\Omega_3\, \Omega_1}{x}\,,\quad
\Omega_3' = -\fft{\Omega_1\, \Omega_2}{1-x}\,,\label{omfo}
\ee
%%%%%
where a prime denotes a derivative with respect to $x$, and $F$ is given by
%%%%%
\be
F= -\fft{8x\, \Omega_1^2\, \Omega_2^2\, \Omega_3^2 + 2\Omega_1\,
\Omega_2\, \Omega_3\, [x\, (\Omega_1^2+\Omega_2^2) -(1-4\Omega_3^2)(
\Omega_2^2 -(1-x)\,\Omega_1^2)]}{4(x\, \Omega_1\, \Omega_2 + 2
\Omega_3\, (\Omega_2^2 - (1-x)\, \Omega_1^2))^2}\,.\label{Fdef}
\ee
%%%%%
(We have normalised the Einstein constant so that $R_{ab}=3g_{ab}$.)

   This first-order system can be reduced to the problem of solving
the Painlev\'e VI equation \cite{tod}.  One introduces a function
$y(x)$, in terms of which the $\Omega_i^2$ are written as
%%%%%
\bea
\Omega_1^2 &=& \fft{(y-x)^2\, y\, (y-1)}{x\, (1-x)}\, \Big(
z-\fft1{2(y-1)}\Big) \, \Big(z-\fft1{2y}\Big)\,,\nn\\
\Omega_2^2 &=& \fft{y^2\, (y-1)(y-x)}{x}\,  \Big(
z-\fft1{2(y-x)}\Big) \,\Big( z-\fft1{2(y-1)}\Big)\,,\\
\Omega_3^2 &=& \fft{(y-1)^2\, y\, (y-x)}{(1-x)}\,  \Big(
z-\fft1{2y}\Big) \,\Big( z-\fft1{2(y-x)}\Big)\,,\nn
\eea
%%%%%
where
%%%%%
\be
z = \fft{x-2x\, y + y^2 -2x\, (1-x)\, y'}{4y\, (y-1)(y-x)}\,.
\ee
%%%%%
(Note that $\Omega_1^2-\Omega_2^2 -\Omega_3^2$, which is conserved,
must take the value $-\ft14$ in order that $ds_4^2$ be Einstein.)
The claim then is that
the first-order equations are satisfied if $y$ satisfies the 
Painlev\'e VI equation
%%%%%
\bea
y''&=& \ft12\Big(\fft1{y} + \fft1{y-1} + \fft1{y-x}\Big)\, {y'}^2
- \Big(\fft1{x} + \fft1{x-1} + \fft1{y-x}\Big)\, y'\nn\\
&& + \fft{y\,(y-1)(y-x)}{x^2\, (x-1)^2}\, \Big(\a + \beta\, 
\fft{x}{y^2} + \gamma\, \fft{x-1}{(y-1)^2} + \delta\, 
 \fft{x\, (x-1)}{(y-x)^2}\Big)\,,\label{painleve}
\eea
%%%%%
with $(\a,\beta,\gamma,\delta)=(\ft18, -\ft18, \ft18,\ft38)$.
Note that the expression (\ref{Fdef}) for $F$ is actually quite
simple, expressed in terms of $y$:
%%%%%
\bea
F &=&\fft{y\, (1-y)\, (y-x)\, z}{2x\, (1-x)}\nn\\
&=& \fft{x-2x\, y + y^2 - 2x\, (1-x)\, y'}{8x\, (1-x)}\,.
\eea
%%%%%

   It is a straightforward, although somewhat involved, exercise to
show that if the first-order equations (\ref{omfo}) are satisfied,
then the metric functions $a_i$ indeed satisfy our first-order
equations (\ref{foeq}).  Note, however, that the converse is not true;
not every solution of the general first-order equations (\ref{foeq})
for anti-self-dual Einstein metrics gives a solution of (\ref{omfo}).
For example, the uniaxial solutions certainly do not satisfy the
equations (\ref{omfo}); setting the $a_i^2$ equal implies that
$\Omega_2^2=(1-x)\, \Omega_1^2$ and $\Omega_3^2= x\, \Omega_1^2$, and
one can easily see that substituting into (\ref{omfo}) leads to a
contradiction.  Likewise, one can show that setting any two of the
metric functions equal leads to a degeneration in (\ref{omfo}).  This
can be understood from the fact that the radial coordinate used in
\cite{tod,hitchin} becomes a constant if any two of the metric
functions are set equal.

   The first-order equations (\ref{omfo}) were obtained in
\cite{tod,hitchin,hitchin2} by first solving the conditions for
metrics with anti-self-dual Weyl tensor and vanishing Ricci scalar,
and then performing a conformal rescaling of the metric to arrive at
one that was Einstein.  We have shown that every solution of the
Tod-Hitchin system provides a solution of our system of first-order
equations.  Our equations are valid not only for the triaxial case but
also for the biaxial and uniaxial cases, and yield all possible
Bianchi IX self-dual Einstein metrics.  The method of Tod and Hitchin
breaks down in the biaxial and uniaxial cases.  The arguments from
twistor theory presented in \cite{hitchin2} show that the Tod-Hitchin
method gives the general triaxial metric, but the explicit
correspondence to our first-order equations remains unclear.

\subsection{Explicit examples} \label{hitchinexplicit}

   Hitchin gives explicit solutions to (\ref{painleve}) characterised
by an integer $k$, with $k=3,4,6,8$ \cite{hitchin,hitchin2}.  The case
$k=3$ corresponds to the round metric on $S^4$, written in triaxial
form \cite{giulini}, whilst $k=4$ corresponds to the Fubini-Study
metric on $\CP^2$, again written in triaxial form
\cite{bougib}.\footnote{The triaxial form of the Fubini-Study metric
is derived in appendix \ref{dragtsec}.}  For $k\ge5$ the metrics will
necessarily have orbifold-type singularities.

    In general it is easiest to give these solutions $y(x)$ by
introducing a ``parametric variable'' $r$, with $y$ and $x$ both
expressed in terms of $r$.  Thus one has:
%%%%%
\bea
k=3:&& y=\fft{r^2\, (2r^2+5r+2)}{(2r+1)(r^2+r+1)}\,,\qquad
x= \fft{r^3\, (r+2)}{2r+1}\,,\nn\\
k=4:&& y= r\,,\qquad x=r^2\,,\nn\\
k=6:&& y= \fft{r\, (r^2+r+1)}{(2r+1)}\,,\qquad
x= \fft{r^3\, (r+2)}{(2r+1)}\,,\nn\\
k=8:&& \fft{4r\, (3r^2-2r+1)}{(r+1)(1-r)^3\, (r^2+2r+3)}\,,\qquad
x=\Big(\fft{2r}{1-r^2}\Big)^4\,.\label{parsol}
\eea
%%%%%
It is straightforward to verify that these expressions all satisfy the
Painlev\'e equation (\ref{painleve}).

   For $k=3$, after normalising so that $R_{ab}=3 g_{ab}$, the
metric (\ref{ommetric}) becomes \cite{hitchin}
%%%%%
\be
h^2= \fft{3}{(1+r+r^2)^2}\,,\quad a_1^2=(1+2r)^2\, h^2\,,
\quad a_2^2= (1-r^2)^2\, h^2\,,\quad a_3^2 = r^2\, (2+r)^2\, h^2\,.
\ee
%%%%%
Note that the radial variable $r$ being used here is precisely the 
parametric variable in (\ref{parsol}).
Defining a new radial variable $t$ by $r=-\ft12 + \ft{\sqrt3}{2}\,
\tan(\ft12\sqrt3\,t)$, the $k=3$ metric becomes
%%%%%
\be
ds_4^2 = dt^2 + 4\sin^2 t \, \sigma_1^2 + 
4\sin^2(t-\ft23\pi)\, \sigma_2^2 + 4\sin^2(t+\ft23\pi)\, \sigma_3^2\,,
\label{s4triaxial}
\ee
%%%%%
which can be recognised as the triaxial form of the Einstein metric on
$S^4$, discussed in \cite{giulini}.

   For $k=4$, and normalising for convenience so that  $R_{ab}=6 g_{ab}$,
the metric in \cite{hitchin} has
%%%%%
\be
h^2=\fft1{4r\, (1+r)^2}\,,\quad a_1^2= \fft{1}{1+r}\,,\quad
a_2^2=\fft{(1-r)^2}{(1+r)^2}\,,\quad
a_3^2= \fft{r}{1+r}\,.
\ee
%%%%%
Defining a new radial variable by $r=\tan^2 t$, the $k=4$
metric becomes
%%%%%
\be
ds_4^2  = dt^2 + \cos^2 t\, \sigma_1^2 +
\cos^2 2t\, \sigma_2^2 +
\sin^2 t\, \sigma_3^2\,,\label{cp2triaxial}
\ee
%%%%%
which can be recognised as the triaxial $\CP^2$ metric \cite{bougib}, 
discussed in appendix \ref{dragtsec}.

  For $k=6$, the metric functions are given by
%%%%%
\bea
&&h^2 = \fft{3(1+r+r^2)}{r\, (r+2)^2\, (2r+1)^2}\,,\qquad
a_1^2 = \fft{3 (1+r+r^2) }{(r+2)\, (2r+1)^2}\,,\nn\\
&&a_2^2 = \fft{3(r^2-1)^2}{(1+r+r^2)\, (r+2)\, (2r+1)}\,,\qquad
a_3^2 = \fft{3r\, (1+r+r^2)}{(r+2)^2\, (2r+1)}\,.\label{k6metric}
\eea
%%%%%
The radial coordinate runs from $r=1$ to $r=\infty$, and we have normalised 
the metric so that $R_{ab}= 3 g_{ab}$.  

   For $k=8$, after rederiving the metric using the construction given
in \cite{hitchin}, we find that the metric functions are given by
%%%%%
\bea
h^2 &=& \fft{4(1+r)(3-2r+r^2)(1-2r+3r^2)(1+2r+3r^2)}{
   (1-r)\, r\, (1+r^2)(1+2r-r^2)^2\, (3+2r+r^2)^2}\,,\nn\\
a_1^2 &=& \fft{4(1-r)(1+r)^3\, (3-2r+r^2)(1-2r+3r^2)}{
         (1+2r-r^2)(3+2r+r^2)^2\, (1+2r+3r^2)}\,,\nn\\
a_2^2 &=& \fft{4(1+r^2)(3-2r+r^2)(1-2r-r^2)^2\, (1+2r+3r^2)}{
  (1+2r-r^2)^2\, (3+2r+r^2)^2\, (1-2r+3r^2)}\,,\nn\\
a_3^2 &=& \fft{16r\, (1-2r+3r^2)(1+2r+3r^2)}{(1+2r-r^2)(3-2r+r^2)
(3+2r+r^2)^2}\,,\label{k8metric}
\eea
%%%%%
where we have again chosen the normalisation so that $R_{ab}=3g_{ab}$.
(This corrects a typographical error in \cite{hitchin}, where there is
an extra factor $(1+r)^2$ in the coefficient of $\sigma_3^2$ that
should not be there.)  The radial coordinate lies in the interval
$\sqrt2-1 < r <1$.

    The $k=3$ and $k=4$ Tod-Hitchin metrics are $S^4$ and $\CP^2$
respectively, albeit in their less common triaxial forms.  The
existence of more than one Bianchi IX form is a consequence of the
homogeneity of these metrics.  The isometry algebra contains more than
one $SU(2)$ subalgebra, and the orbits are different.  The full set of
homogeneous Einstein 4-manifolds is known, and from that list we
deduce that this can only happen for self-dual Einstein metrics in the
case of $S^4$ and $\CP^2$.  Thus for higher values of $k$, the Tod-Hitchin
metrics and the biaxial Bianchi IX self-dual Einstein metrics form 
disjoint classes.  An explicit demonstration of this for the 
$k=6$ and $k=8$ metrics can be given by computing the quantity $X^3/Y^2$,
where $X$ and $Y$ are the quadratic and cubic Weyl tensor invariants 
defined in (\ref{xydef}).  We showed that any biaxial self-dual Einstein
metric must satisfy $X^3/Y^2=1$ (see (\ref{xyrel})), and an elementary
calculation shows that whilst this is true for the $k=3$ and $k=4$
metrics, it does not hold for the $k=6$ and $k=8$ metrics.

\subsection{Global structure of the metrics}\label{globalsec}

   The global structure of the Tod-Hitchin metrics is described in 
detail in \cite{hitchin,hitchin2}.  Here, we summarise the conclusions, 
presenting them in a way that is perhaps more readily accessible
to physicists.

   The key to understanding the global structure is to understand the
nature of the degenerate orbits where metric coefficients vanish.
An important feature of the metrics, for all $k$ including $k=3$
and $k=4$, is that at one end of the radial coordinate range the
coefficient of $\sigma_1$ vanishes, while at the other end it is
the coefficient of $\sigma_2$ that vanishes instead. This ``slumping''
is reminiscent of the metric behaviour in the Atiyah-Hitchin metric,
where the coefficient of one of the $\sigma_i$ vanishes at short
distance, while the coefficient of another of them stablises in the
asymptotic region.  In fact, as shown in \cite{hitchin}, the
Atiyah-Hitchin metric itself arises as the $k\longrightarrow \infty$
limit of the Tod-Hitchin metrics.

   Because of the slumping, it is useful to introduce two different
Euler-angle parameterisations of the left-invariant 1-forms, one
adapted to the region where $\sigma_1$ collapses, and the other
adapted to the region where $\sigma_2$ collapses.  The procedure was
described in \cite{gibman}, and elaborated somewhat in
\cite{cglpslump}. Here we shall present a brief summary of the 
description in \cite{cglpslump}, with labelling adapted to our present 
conventions.  

    Let us introduce Euler angles $(\theta,\phi,\psi)$ and
$(\td\theta,\td\phi,\wtd\psi)$, such that
%%%%%
\bea
&&\sigma_1=d\psi+\cos\theta\, d\phi\,,\quad
\sigma_2+\im\, \sigma_3= e^{\im\, \psi}\, (d\theta+\im\, \sin\theta\, 
d\phi)\,,\label{sig1}\\
&&\sigma_2=d\wtd\psi+\cos\td\theta\, d\td\phi\,,\quad
\sigma_3+\im\, \sigma_1= e^{\im\, \wtd\psi}\, (d\td\theta+\im\, 
\sin\td\theta\, d\td\phi)\,.\label{sig2}
\eea
%%%%%
We begin by taking $\psi$ and $\wtd\psi$ both to have period $2\pi$, so 
that the orbits are $\RP^3$.  
Clearly one could, in principle, solve for the transformation that
relates the tilded and untilded coordinates, but we shall not need this.

   We now consider the operation, which we shall denote by $I_1$,
which implements the identification $\psi\approx \psi+\pi$.  It is
easily seen that in terms of the tilded coordinates, this corresponds
to $\td\theta\longrightarrow \pi-\td\theta$, $\td\phi\longrightarrow
\td\phi +\pi$, $\wtd\psi\longrightarrow -\wtd\psi$.  Likewise we 
define $\wtd I_2$ which implements $\wtd\psi\approx \wtd\psi+\pi$.  Since the
tilded basis is related to the untilded by a cyclic permutation of 
$(\sigma_1,\sigma_2,\sigma_3)$, we can see that in our notation we shall 
have $I_i=\wtd I_i$, and so we can deduce that the effect of the $I_i$ 
on the untilded coordinates is
%%%%%
\bea
I_1:&& \theta\longrightarrow \theta\,,\qquad 
    \phi\longrightarrow\phi\,,\qquad
   \psi\longrightarrow \psi+\pi\,,\nn\\
I_2:&& \theta\longrightarrow \pi-\theta\,,\qquad
   \phi\longrightarrow \phi+\pi\,,\qquad
  \psi\longrightarrow -\psi\,,\label{untilded}\\
I_3:&& \theta\longrightarrow \pi-\theta\,,\qquad
   \phi\longrightarrow \phi+\pi\,,\qquad
  \psi\longrightarrow \pi -\psi\,,\nn
\eea
%%%%%
while on the tilded coordinates we have
%%%%%
\bea
I_1:&& \td\theta\longrightarrow \pi-\td\theta\,,\qquad
   \td\phi\longrightarrow \td\phi+\pi\,,\qquad
  \wtd\psi\longrightarrow \pi -\wtd\psi\,,\nn\\
I_2:&& \td\theta\longrightarrow \td\theta\,,\qquad 
    \td\phi\longrightarrow\td\phi\,,\qquad
   \wtd\psi\longrightarrow \wtd\psi+\pi\,,\label{tilded}\\
I_3:&& \td\theta\longrightarrow \pi-\td\theta\,,\qquad
   \td\phi\longrightarrow \td\phi+\pi\,,\qquad
  \wtd\psi\longrightarrow -\wtd\psi\,,\nn
\eea
%%%%%   
   
   Consider first the case $k=3$, which gives the triaxial metric 
(\ref{s4triaxial}) on $S^4$.  Near $t=0$ we have
%%%%%
\be
ds_4^2\sim dt^2 + 4 t^2\, \sigma_1^2 + \sigma_2^2+\sigma_3^2\,.
\ee
%%%%%
From the expression (\ref{sig1}) we see that regularity at $t=0$
requires that $\psi$ have period $\pi$, and so from (\ref{untilded})
we should impose the identification $I_1$.  Near the other endpoint
$t=\ft23\pi$, we set $t=\ft23 \pi- \tau$, and so the metric takes the
form
%%%%%
\be
ds_4^2 \sim d\tau^2 + 4\tau^2\, \sigma_2^2 + \sigma_1^2 + \sigma_3^2\,.
\ee
%%%%
From (\ref{sig2}) we see that regularity requires that $\wtd\psi$ have
period $\pi$, and so from (\ref{tilded}) we should in addition impose the
identification $I_2$.  Thus the principal orbits are $SO(3)/(\bZ_2\times 
\bZ_2)$. We also see that the 2-dimensional bolt described by 
$\sigma_2^2 +\sigma_3^2=d\theta^2 + \sin^2\theta\, d\phi^2$ at $t=0$, 
and the 2-dimensional bolt described by $\sigma_1^2 +\sigma_3^2 =
d\td\theta^2 + \sin^2\td\theta\, d\td\phi^2$ at $t=\ft23\pi$ each has 
the topology of $\RP^2$, since there is an antipodal identification 
on the former implied by $I_2$ in (\ref{untilded}), and in the latter 
implied by $I_1$ in (\ref{tilded}).  The metric therefore extends smoothly
on the Veronese surfaces $\RP^2$ at each endpoint \cite{hitchin}.

   The case $k=4$ gives the triaxial $\CP^2$ metric (\ref{cp2triaxial}).
We can take the two endpoints to be at $t=\ft12\pi$ and $t=\ft14\pi$.
Near $t=\ft12\pi$, after setting $t=\ft12\pi-\tau$ the metric takes the form
%%%%%
\be
ds_4^2 \sim d\tau^2 + \tau^2\, \sigma_1^2 + \sigma_2^2 + \sigma_3^2\,.
\ee
%%%%%
Regularity therefore requires that we {\it not} impose the identification
$I_1$.  On the other hand, at the other endpoint $t=\ft14\pi$, after 
defining $t=\ft14\pi-\tau$ we have
%%%%%
\be
ds_4^2 \sim d\tau^2 + 4\tau^2\, \sigma_2^2 + \sigma_1^2+\sigma_3^2\,.
\ee
%%%%% 
Regularity therefore requires that we impose the identification $I_2$. 
This means that the principal orbits are $SO(3)/\bZ_2$, and that the
metric extends smoothly onto $\RP^2$ at $t=\ft12\pi$, and onto $S^2$ 
at $t=\ft14\pi$ \cite{hitchin}.  This reflects the fact that $\CP^2$ 
can be described as the double covering of $S^4$ branched over $\RP^2$. 

   For $k=6$, we see by letting
$r=1+3\rho$ that near $r=1$ the metric (\ref{k6metric}) takes the form
%%%%%
\be
ds_4^2 \sim d\rho^2 + 4 \rho^2\, \sigma_2^2 + \ft13(\sigma_1^2
+\sigma_3^2)\,,
\ee
%%%%%
whilst letting $r=1/\rho^2$ the metric near $r=\infty$ has the form
%%%%%
\be
ds_4^2 \sim d\rho^2 + \ft14\rho^2\, \sigma_1^2 + \ft32(\sigma_2^2 
+\sigma_3^2)\,.
\ee
%%%%%
Thus if we impose the identification $I_2$ the metric extends smoothly
over $\RP^2$ at $r=1$, and extends over $\RP^2$ with an orbifold
singularity having angle $\ft12\pi$ at $r=\infty$ \cite{hitchin}.  

    For $k=8$, after letting $r=\sqrt2-1+ (\sqrt{2-\sqrt2})\, \rho$,
the metric near $r=\sqrt2-1$ can be seen to have the form
%%%%%
\be
ds_4^2 \sim d\rho^2 + 4 \rho^2\, \sigma_2^2 + (3-2\sqrt2)\, (\sigma_1^2
+\sigma_3^2)\,.
\ee
%%%%%
Letting $r=1-\ft38 \rho^2$, the metric near $r=1$ takes the form
%%%%%
\be
ds_4^2 \sim d\rho^2 + \ft19 \rho^2\, \sigma_1^2 + \ft43(
\sigma_2^2 + \sigma_3^2)\,.
\ee
%%%%%
Thus by imposing the identification $I_2$ the metric extends smoothly
over $\RP^2$ at $r=\sqrt2-1$, and extends over $\RP^2$ with an
orbifold singularity having angle $\ft13\pi$ at $r=1$ \cite{hitchin}.
 
   In \cite{hitchin} it is shown that all the metrics obtained from
solving the Painlev\'e equation are positive definite with $x$ lying
in the interval $1 <x < \infty$, for all values of the constant $k$
parameterising the solutions described in \cite{hitchin}.  Near $x=1$,
the metric takes the form
%%%%%
\be
ds_4^2 \sim \ft1{16} \cos^2\fft{\pi}{k}\, (dr^2 + 4 r^2\, \sigma_2^2) 
+\sigma_1^2 + \sigma_3^2\,.
\ee
%%%%%
This shows that the metric extends over the degenerate orbit at $r=0$, 
with $\sigma_1^2 +\sigma_3^2$ describing $\RP^2$ \cite{hitchin}.  As
$x\longrightarrow\infty$ the metric assumes the form
%%%%%
\be
ds_4^2 \sim d\rho^2 + \fft{4\rho^2}{(k-2)^2}\, \sigma_1^2 + 
  2^{8/k-2}\, (\sigma_2^2 + \sigma_3^2)\,,
\ee
%%%%%
where $x=\rho^{-k}$, which shows that there is an orbifold singularity
with angle $2\pi/(k-2)$ around $\RP^2$ \cite{hitchin}.  These 
results are consistent with the explicit calculations for the $k=6$ 
and $k=8$ cases above.

\section{Singularity structure and M-theory}

   In this paper we have extended the analysis of $G_2$ holonomy
spaces to those whose principal orbits are twistor spaces, constructed
as $S^2$ bundles over four-dimensional self-dual Einstein metrics of
the general Bianchi IX type. We obtained the general first-order
differential equations for these triaxial Bianchi IX metrics, 
and we showed how they can be derived from a superpotential.
In special cases, the self-dual Einstein metrics reduce to
$S^4$, $\CP^2$ and the (biaxial) Taub-NUT-de Sitter metrics, 
    
   We focused on the analysis of the local and global structures of
the self-dual Einstein Bianchi IX metrics. For the biaxial
specialisation, where the local form of the general solution is well
known, we gave a complete analysis of the solutions by studying the
flows in the phase-plane of the first-order equations.  Even in this 
biaxial case the analysis is quite subtle, since there is no single
local expression for the metric that directly covers all the possible 
regions of flows in the phase-plane.  Some regions are well-described
by the standard expression for the self-dual Taub-NUT-de Sitter metrics,
but our analysis reveals that in another region there are flows that
are more appropriately described by a different local form of the solution, 
which we refer to as the self-dual Eguchi-Hanson-de Sitter metrics.
These metrics, which as far as we are aware have not been presented 
explicitly before, describe flows in a region of the phase-plane 
that can be viewed as generalisations of the Eguchi-Hanson metric in 
which the cosmological constant is non-zero.  Unlike the usual 
Eguchi-Hanson-de Sitter metrics \cite{gibpop}, which are K\"ahler but neither
self-dual nor anti-self-dual, the new metrics have a self-dual Weyl tensor
even when the cosmological constant is non-zero.  In the self-dual
Taub-NUT-de Sitter form, the two parameters of biaxial solutions can
be thought of as the NUT parameter and the cosmological constant.  In the
self-dual Eguchi-Hanson-de Sitter form, the two parameters can be thought of
as the Eguchi-Hanson scale size and the cosmological constant.

    We discussed the global structure for the biaxial self-dual
metrics, both for positive and negative cosmological constant.  For
the positive cosmological constant the metrics are compact, in general
with singularities.  The radial coordinate ranges over an interval
that terminates at endpoints where the $SU(2)$ principal orbits
degenerate; to a point (a NUT) at one end, and to a two-dimensional
surface (a bolt) that is (locally) $S^2$ at the other.  For generic
choices of the NUT parameter (or, in the alternative local
description, the Eguchi-Hanson scale size), the metrics cannot be
smoothly extended on the NUT and bolt endpoints simultaneously.  This
is because the periodicity requirements needed for regularity at one
end are in general incommensurate with the periodicity requirements at
the other end.  Only for very special values of the NUT parameter
is the metric regular at both endpoints. In general, however, one
encounters singularities at either endpoint of the four-dimensional
radial coordinate.

    In the generic case, a specific choice of the period for the
azimuthal angle $\psi$ allows the singularity at the $S^2$ bolt to be
removed, but then the NUT has a co-dimension four orbifold
singularity. Alternatively, choosing the periodicity appropriate for
regularity at the NUT, there will be a co-dimension two singularity on
the $S^2$ bolt.  The associated seven-dimensional $G_2$ holonomy space
therefore has singularities of the same co-dimensions. The
co-dimension four NUT singularities may admit an M-theory
interpretation associated with the appearance of non-abelian gauge
symmetries \cite{behrndt} and the circle reduction of M-theory on
these $G_2$ holonomy spaces may have a Type IIA interpretation in
terms of a location of coincident D6-branes \cite{behrndt}.  On the
other hand the co-dimension two singularities at the bolts do not seem
to have a straightforward interpretation in M-theory dynamics.  Since
neither of type of singularity is of co-dimension seven, these spaces
do not seem to shed light on the appearance of chiral matter.

    The triaxial self-dual Einstein Bianchi IX metrics described by
the Tod-Hitchin system are defined on compact spaces with bolts at
each endpoint. For the solutions discussed in section
\ref{hitchinexplicit}, with $k\ge 6$, one endpoint has an $\RP^2$ bolt,
while the other endpoint is a $\RP^2$ bolt with a $\Z_{k-2}$ conical
co-dimension two singularity. The corresponding $G_2$ holonomy spaces
again have co-dimension two singularities, and so M-theory on these
spaces does not have a straightforward interpretation; in particular
their relevance for obtaining non-abelian gauge group enhancement or
the appearance of chiral matter is not clear.
   
   Despite the fact that the role of the singularities in our metrics
in M-theory is unclear, one thing is certain: the singularities do not
affect the amount of supersymmetry.  Because the Killing spinor is 
a singlet, it is invariant under all elements of the isometry group.
In particular, it is invariant under the action of the binary 
dihedral group generated by $I_1$, $I_2$ and $I_3$, and in the biaxial
case it is invariant under arbitrary shifts of the coordinate $\psi$. 
Since it was these symmetries that entered into the discussion of 
singularities, it is clear that no matter what identifications we
choose to make, it will not affect the existence of the Killing
spinor.   This should be contrasted with the co-dimension two and 
co-dimension four singularities discussed in \cite{panic}.  In that
case, the Killing spinors are not singlets, and identifications may or
may not leave them invariant.  The singularities for which
the identifications are incompatible with the existence of Killing
spinors are believed to be unstable, due to closed-string tachyons,
whilst those that are compatible with the Killing spinors are believed
to be stable.  In our case, it is clear that there is no room for
a closed-string tachyon instability, or its M-theoretic analogue.
In other words, ``Don't Panic, it's $G_2$!''

\section*{Acknowledgements}

   We should like to thank Klaus Behrndt, Nigel Hitchin and Paul Tod
for useful discussions.  M.C and C.N.P. are grateful to the Michigan
Center for Theoretical Physics and the Isaac Newton Institute for
Mathematical Sciences for hospitality and financial support during the
course of this work.  M.C.~is supported in part by DOE grant
DE-FG02-95ER40893 and NATO grant 976951; H.L.~is supported in full by
DOE grant DE-FG02-95ER40899; C.N.P.~is supported in part by DOE
DE-FG03-95ER40917.  G.W.G.~acknowledges partial support from PPARC
through SPG\#613.

\newpage
   
\centerline{\Large\bf APPENDICES}

\appendix
\section{Bianchi IX Einstein-K\"ahler metrics}\label{ekmetsec}

    The purpose of this appendix is to clarify the distinction between
the anti-self-dual Einstein metrics considered in this paper and Bianchi IX
Einstein-K\"ahler metrics.  These two classes do not overlap except
when the metrics are Ricci-flat, or else the Fubini-Study metric on
$\CP^2$ (or the Bergmann metric on the open ball in $\C^2$ if
$\Lambda<0$).  In the case that the metrics are biaxial, the general
Einstein-K\"ahler solutions, together with their K\"ahler potential,
were obtained in \cite{gibpop}, where they were called the
Eguchi-Hanson-de Sitter metrics (see equation (\ref{ehds})).  A subsequent
discussion was given in \cite{pedersen}.

   The triaxial case has been considered by Dancer and Strachan in
\cite{danstr}, where a first-order system was obtained.  This
generalises that for hyper-K\"ahler metrics with triholomorphic
$SU(2)$ action, written down and solved in \cite{begipapo}.  The
general solution of the Dancer-Strachan system is not known, but
particular cases, such as triaxial forms of the Fubini-Study metric 
on $\CP^2$ and the product metric on $\CP^1\times \CP^1$ are
known, and turn out to be remarkably simple.  

   Writing the Bianchi IX metrics in the form (\ref{bianchi9}), with
$e^0=dt$ and $e^i=a_i\, \sigma_i$, a basis for anti-self-dual 2-forms is
$\Omega_i=e^0\wedge e^i - \ft12 \ep_{ijk}\, \e^j\wedge e^k$, and so an
ansatz for the $SU(2)$-invariant anti-self-dual K\"ahler form is
%%%%%
\be
\Omega =\a_i\, \Omega_i\,,
\ee
%%%%%
where the coefficients $\a_i$ depend only on $t$, and $\a_i^2=1$.  
The metric will be K\"ahler if $\Omega$ is covariantly constant, which leads 
to the first-order equations
%%%%%
\be
\dot \a_1 = (\beta_3+\gamma_3)\, \a_2 -(\beta_2+\gamma_2)\, \a_3\,,\qquad
\etcyc\,,
\ee
%%%%%
where $\beta_i$ and $\gamma_i$ are defined in (\ref{bega}).  From
these, and the Einstein equations, one can show that $\a_1=\a_2=0$ and
$\a_3=1$ (or cyclic permutations) \cite{danstr}, and hence that the
metric coefficients satisfy the first-order equations
%%%%%
\bea
\dot a_1 &=& -\fft{a_1^2 -a_2^2-a_3^2}{2 a_2\, a_3} \,,\nn\\
\dot a_2 &=& -\fft{a_2^2 -a_1^2-a_3^2}{2 a_1\, a_3} \,,\label{ekfo}\\
\dot a_3 &=& -\fft{a_3^2 -a_1^2-a_2^2 +2\Lambda\, 
                   a_1^2\, a_2^2}{2 a_1\, a_2}  \,,\nn
\eea
%%%%%

   Rewriting in terms of the radial variable $\eta$, defined by
$dt=a_1\, a_2\, a_3\, d\eta$, it is easily seen that the first-order
equations can be derived from a superpotential.  In the notation of
section \ref{superpotsec}, the potential $U$ in (\ref{kinpot}) can be
written as $U=-\ft12 g^{ij}\, (\del W/\del \a^i)\, (\del W/\del
\a^j)$, where we now define $\a^i=(\log a_1,\log a_2,\log a_3)$, and hence
$g_{ij}=2-2\delta_{ij}$.  We find that the superpotential is then given by
%%%%%
\be
W= -(a_1^2 + a_2^2 +a_3^2) +\Lambda\, a_1^2 \, a_2^2\,.
\ee
%%%%%

   Two particular triaxial solutions of the first-order Einstein-K\"ahler
system (\ref{ekfo}) are the Fubini-Study metric on $\CP^2$, which can 
be written (setting $\Lambda=6$ for convenience) as \cite{bougib}
%%%%%
\be
ds_4^2 = dt^2 + \sin^2 t\, \sigma_1^2 + \cos^2 t\, \sigma_2^2 
+\cos^2 2t\, \sigma_3^2 \,,\label{triaxialcp2}
\ee
%%%%%
and the product metric on $S^2\times S^2$, which can be written
(setting $\Lambda=2$ for convenience) as \cite{popeth}
%%%%%
\be
ds_4^2 = dt^2 + \sin^2 t\, \sigma_1^2 + \sigma_2^2 + \cos^2 t\, \sigma_3^2
\,.\label{triaxials2s2}
\ee
%%%%%
In view of the somewhat unfamiliar forms of these metrics, we shall give
a brief description of them below.

\section{Iwai's construction, Dragt coordinates and the
Guichardet connection}\label{dragtsec}

    In this appendix, we shall derive the triaxial forms of the
Einstein metrics on $\CP^2$ and $S^2\times S^2$.  The method used 
differs slightly from the ones in \cite{bougib} and \cite{popeth}, but
it has the merit of giving a unified description of the two cases.
The basic idea is to express the metric in flat Euclidean 6-space in
an appropriate coordinate system, adapted to an $SO(3)$ action.  We
shall here follow the paper of Iwai \cite{iwai}, who was interested in
the three-body problem in molecular physics.  It turns out that we can
use his results not only to obtain Bianchi IX metrics but we can also use
Scherk-Schwarz reduction to obtain some insight into global monopoles
of the sort recently studied by Hartnoll \cite{hart}.

   We think of $\bE^6$ as $\bE^3 \oplus \bE^3 \ni ({\bf x}, {\bf y})$
and consider the diagonal action\footnote{Note that the triaxial form
of the standard round metric on $S^4$ can also be obtained from the 
flat metric on $\bE^6$, but now the action of $SO(3)$ is different.
In this case one identifies $\bE^6$ with the space of real symmetric
$3\times 3$ matrices on which $SO(3)$ acts by conjugation \cite{giulini}.}
of $SO(3)$. Projection from the
principal orbits is Iwai's generalisation of the standard Hopf map
used in the Taub-NUT metric. This standard Hopf map $\pi: \bE^4 \equiv
(\bC \oplus \bC) \ni (z^1, z^2) \rightarrow \bR^3
\equiv \bR \oplus \bC$ onto the orbits of the diagonal action
of $U(1)$ given by
%%%%%
\be
(z^1, z^2) \rightarrow (|z^1|^2-|z^2|^2, 2 z^1 {\bar z^2} )\,.
\ee
%%%%%
Introducing polar coordinates on $\bR^3$, and an angle $\psi$ along
the Hopf fibres, we may write the flat metric on $\bE^4$ as a special
case of the multi-centre metrics, which have an interpretation in
terms Kaluza-Klein monopoles and D6-branes. Iwai's procedure is rather
similar and may have a corresponding generalisation.

  In the case of flat six dimensions, Iwai's map is $\pi: \bE^6
\rightarrow \bR^3 _+ \ni ( w^1, w^2, w^3)$, given by
%%%%%
\be
({\bf x}, {\bf y} ) \rightarrow ( {\bf x} ^2 - {\bf y}^2, 2 {\bf
x} \cdot {\bf y} , 2 | {\bf x} \times {\bf y} | ) = ( w^1, w^2 ,
w^3)\,,
\ee
%%%%%
with $w^3 \ge 0$.  Th orbit space $\bR^3_+$ may be given coordinates
$(\rho, \psi ,\chi)$, called Dragt coordinates, such that
%%%%%
\be
(w^1, w^2, w^3)= ( \rho ^2 \cos \psi \cos \chi , \rho^2 \sin \psi
\cos \chi   , \rho ^2 \sin \chi)\,,
\ee
%%%%%
with $0\le \rho < \infty$,  $0 \le \psi <2 \pi$, $0 \le \chi <
{\pi \over 2} $. Note the range of $\chi$.  One checks that
%%%%%
\be
{\bf x} ^2 +{ \bf y} ^2 = \rho ^2 = \sqrt { (w^1) ^2 + ( w^2 ) ^2
+ ( w^3) ^2 }\,.
\ee
%%%%%
To fix the $SO(3)$ freedom we introduce an orthonormal moving
frame $({\bf u}_1, {\bf u}_2,  {\bf u} _3)$ related to a fixed
orthonormal frame $({\bf e} _1 ,{\bf e} _2 ,{\bf e}_3) $ by a
rotation with standard Euler angles and left-invariant 1-forms
$(\sigma _1, \sigma _2 ,\sigma _3)$ say. Now if
%%%%%
\be
{\bf x} =\rho \cos{\psi \over 2} \cos {\chi \over 2} {\bf u} _1 -
\rho \sin {\psi \over 2} \sin{ \chi \over 2} {\bf u}_2 \,,
\ee
%%%%% 
and
%%%%%
\be
{\bf y} = \rho \sin { \psi \over 2} \cos {\chi \over 2} {\bf u} _1
+ \rho \cos {\psi \over 2} \sin {\chi \over 2} {\bf u} _2 \,,
\ee
%%%%%
Iwai finds that the flat metric on $\bE^6$ is given by
%%%%%
\be
ds^2 = d \rho ^2 + \ft14 \rho ^2 ( d \chi ^2 + \cos ^2 \chi
d \psi ^2 ) + \rho ^2 \sin ^2 {\chi \over 2} \,\sigma _1^2 + \rho ^2
\cos ^2 {\chi \over 2} \, \sigma _2 ^2 + \rho ^2 ( \sigma _3 - \ft12 
\sin \chi \, d\psi ) ^2 \,.
\ee
%%%%%

   If we set $\psi = \ft12 \pi$ and $\rho^2 =2$, the vectors ${\bf x}$
and ${\bf y}$ have unit magnitudes, and thus parameterise points on
$S^2\times S^2$, embedded in $\bR^3\times \bR^3$.  The result
is the metric (\ref{triaxials2s2}) on $S^2\times S^2$, obtained in
\cite{popeth}.

   If instead we set $\rho =1$ we obtain the unit $S^5$.  The angle
$\psi$ is a coordinate along the Hopf fibres.  Projecting orthogonally
to the Hopf fibres, we obtain the triaxial form (\ref{triaxialcp2}) of
the Fubini-Study metric on $\CP^2$ obtained in \cite{bougib}.

  We note {\it en passant} that we could consider the 
seven-dimensional flat metric on $\bE^{6,1}$ as a trivial solution of
supergravity, and perform a Scherk-Schwarz reduction on the
orbits of $SO(3)$. We get in four dimensions a global monopole coupled to an
$SO(3)$ gauge field $A^i$, $i=1,2,3$, with the Higgs field in the
symmetric tensor (\ie the ${\bf 5}$) representation of $SO(3)$. The
gauge connection coincides with the Guichardet connection, and in the
present case the only non-zero gauge field is
%%%%%
\be
A^3 = - \ft12\sin \chi\, d \psi\,.
\ee
%%%%%
Ignoring the Weyl rescaling, the interpretation is as follows.  One
should think of $\psi$ as an azimuthal angle, \ie a longitude, while
$\chi$ is to be thought of as a latitude. Because $\chi \in [0, {\pi
\over 2} )$, there is a deficit solid angle, and hence a conical
singularity at the origin. Moreover, the metric is not asymptotically
flat.  We have an embedding of an abelian monopole into the
non-abelian gauge group $SO(3)$.  This monopole may be thought of as
sitting at the centre of a global monopole supported by a Higgs field.

\section{Killing spinors}

   Since the seven-dimensional metric constructed from the anti-self-dual
Einstein 4-metric according to (\ref{g2conemet}) has $G_2$ holonomy, it
follows that it admits a covariantly-constant spinor.  It is instructive
to look at how this is related to spinors in the four-dimensional base
space.  To do this, we begin by calculating the Lorentz-covariant
exterior derivative on spinors in seven dimensions in terms of quantities
in the four-dimensional base metric. We adopt a notation where quantities
in seven dimensions carry hats, and so we write (\ref{g2conemet})
as $d\hat s_7^2 =
\rho^{-1}\, (D u_i)^2 + 2 \rho\, ds_4^2$, for which we choose the natural
vielbein basis $\hat e^i= \rho^{-1/2}\, D u_i$, $\hat e^a = \sqrt{2\rho}\,
e^a$. The spinor-covariant exterior derivative is given by $\hat\nabla \equiv
d + \ft14 \hat\omega_{AB}\, \hat\Gamma^{AB}$, and after some calculation
we find that this is given by
%%%%%
\bea
\hat \nabla &=& d + \ft14 \omega_{ab}\, \hat\Gamma^{ab}
  -\ft14 \ep_{ijk}\, A^k\, \hat\Gamma^{ij} +\ft1{16}
\rho^{-3/2}\, u_j\, (\ep_{ijk}\, J^k_{ab}\,\hat\Gamma^{ab}
   - 4 \hat\Gamma^{ij})\, \hat e^i \nn\\
&& +\ft18 \rho^{-3/2}\, u_i\, (\ep_{ijk}\, J^j_{ab}\, \hat\Gamma^{kb}
- 2 \hat\Gamma^{ia})\, \hat e^a\,.\label{covder}
\eea
%%%%%

    The covariantly-constant spinor $\hat\eta$ in the seven-dimensional
$G_2$ metric satisfies $\hat\nabla\, \hat\eta=0$.  It can be seen from
(\ref{covder}) that this spinor is annihilated by the terms involving
the $\R^3$ coordinates $u_i$, and that it is independent of $u_i$.  In
fact in this basis we find that $\hat\eta$ is the spinor
that is determined, up to overall $u_i$-independent scale, by the conditions
%%%%%
\be
\hat\Gamma_{ij} \, \hat\eta = \ft14 \ep_{ijk}\, J^k_{ab}\,
\hat\Gamma_{ab}\, \hat\eta\,.\label{etaeq}
\ee
%%%%%
It then follows from (\ref{covder}) that $\hat\eta$ satisfies
%%%%%
\be
(d+\ft14 \omega_{ab}\, \hat\Gamma^{ab} -\ft14 \ep_{ijk}\, A^k\, 
\hat\Gamma^{ij})\,\hat\eta=0\,.\label{diffcon}
\ee
%%%%%

    Decomposing spinors into the tensor product of spinors in the
four-dimensional base and the $\R^3$ fibres, we choose Dirac matrices
$\hat\Gamma_a=\Gamma_a\otimes \oneone$ and $\hat\Gamma_i =
\Gamma_5\otimes \tau_i$.  
The Pauli matrices $\tau_i$ can be viewed as the generators
of an internal $SU(2)$ isospin, and so (\ref{etaeq}) and (\ref{diffcon})
can be written as
%%%%%
\be
J^i_{ab}\, \Gamma^{ab}\, \eta^\a = 4\im\, (\tau_i)^\a{}_\beta\, \eta^\beta
\,,\qquad
\nabla\, \eta^\a -\ft{\im}4 A^i\, (\tau_i)^\a{}_\beta\, \eta^\beta=0\,.
\label{su2cov}
\ee
%%%%%
The second equation is the condition for the 4-component spinor $\eta^\a$ 
with its isospin doublet index $\a$ to be gauge covariantly constant with
respect to the $SU(2)$ Yang-Mills covariant derivative.

   Using (\ref{etaeq}) we can rewrite (\ref{diffcon}) as
the four-dimensional equation
%%%%%
\be
d\eta^\a + \ft14(\omega_{ab} -\ft12 A^i\, J_{ab})\,\Gamma^{ab}\, \eta^\a=0\,.
\label{4eq1}
\ee
%%%%%
With the Yang-Mills connection taken to be the self-dual part of the
four-dimensional spin connection as in (\ref{sdym}), we therefore find
that (\ref{4eq1}) is nothing but
%%%%%
\be
d\eta^\a + \ft14 \omega^-_{ab}\, \Gamma^{ab}\, \eta^\a=0\,,\label{4eq2}
\ee
%%%%%
where $\omega^-_{ab} \equiv \ft12(\omega_{ab} -\ft12 \ep_{abcd}\,
\omega^{cd})$ is the anti-self-dual part of the spin connection.  In fact
it follows from the conditions (\ref{etaeq}) satisfied by $\hat\eta$ that
$\hat\Gamma_{ab}\,\hat\eta$ is self-dual in the four-dimensional base
space, and hence (\ref{4eq2}) reduces simply to $d\eta^\a=0$.  

   It is interesting to note that in the special case of $\CP^2$,
which does not admit an ordinary sin structure, $\eta^\a$ is a 
generalised spinor (in the terminology of \cite{hawkpope}) that is
charged with respect to the Yang-Mills connection $A^i$.  In this case
the connection is actually $SO(3)$-valued, as opposed to
$SU(2)$-valued, and it is this that serves to compensate for the minus
sign that ordinary spinors would acquire upon parallel propagation 
around a family of curves spanning the bolt in $\CP^2$ \cite{hawkpope}.

\end{document}